\documentclass[lineno]{jfm}
\newif\iffigure
\figuretrue

\usepackage{graphicx}
\usepackage{comment}
\usepackage{newtxtext}
\usepackage{newtxmath}
\usepackage{natbib}
\usepackage{hyperref}
\usepackage{bm}
\hypersetup{
    colorlinks = true,
    urlcolor   = blue,
    citecolor  = black,
}

\newcommand{\RomanNumeralCaps}[1]
\nolinenumbers 

\title{Transforming Butterflies into Graphs: Statistics of Chaotic and Turbulent Systems}

\author{Andre N. Souza\aff{1}
  \corresp{\email{andrenogueirasouza@gmail.com}}
  }

\affiliation{
\aff{1} Massachusetts Institute of Technology, Cambridge, Massachusetts, United States 
}

\begin{document}
\maketitle
\nolinenumbers 

\begin{abstract}
 We formulate a data-driven method for constructing finite volume discretizations of a dynamical system's underlying Continuity / Fokker-Planck equation. A method is employed that allows for flexibility in partitioning state space, generalizes to function spaces, applies to arbitrarily long sequences of time-series data, is robust to noise, and quantifies uncertainty with respect to finite sample effects. After applying the method, one is left with Markov states (cell centers) and a random matrix approximation to the generator. When used in tandem, they emulate the statistics of the underlying system. We apply the method to the Lorenz equations (a three-dimensional ordinary differential equation) and a modified Held-Suarez atmospheric simulation (a Flux-Differencing Discontinuous Galerkin discretization of the compressible Euler equations with gravity and rotation on a thin spherical shell). We show that a coarse discretization captures many essential statistical properties of the system, such as steady state moments, time autocorrelations, and residency times for subsets of state space. 
\end{abstract}

\section{Introduction}
\label{introduction}

Often the goal of modeling a complex system is not to determine the dynamical equations but rather to construct models that converge in distribution to relevant statistics. In the context of turbulence modeling, this can be viewed as one of the goals of a Large Eddy Simulation (LES), where subsets of statistics (often the kinetic energy spectra) are compared to that of Direct Numerical Simulation (DNS). Similarly, in the context of Earth Systems Modeling, the unpredictability of weather patterns over long time scales necessitates the development of nonlinear models that are queried for relevant statistics. Thus the models are not meant to converge to dynamical trajectories but rather, converge in distribution to target observables.

The present work is motivated by the need to construct simplified statistical models of complex physical phenomena such as turbulence. We take on a dynamical systems view of turbulence original to \cite{Hopf1948}, complemented by \cite{Lorenz1963} and found in its modern form in \cite{ChaosBook}. Thus, the approach is to develop a direct discretization of the statistics associated with chaotic or turbulent dynamics, which we assume to be ergodic, mixing, and associated with a fractal manifold.

There exist many types of discretizations that directly target the statistics, which here means a discretization of the underlying continuity equation (deterministic dynamics), Fokker-Planck equation (stochastic dynamics), Perron-Frobenius/Transfer operator (discrete-time dynamics), or Koopman operator (adjoint of the Perron-Frobenius/Transfer operator). Discretizations methods include that of \cite{Ulam1964, Froyland2001, Dellnitz2005} or, for the stochastic Lorenz equations, \cite{Marston2016}. Modern methods take on an operator theoretic plus data-driven approach leading to the construction of Koopman operators that are measure preserving \cite{Colbrook2022}, which build off of earlier work on linearized dynamic operators such as \cite{Schmid2010}. Data-driven construction of the Perron-Frobenius operator is reviewed by \cite{Klus2016} and \cite{Fernex2021}. Convergence guarantees under various assumptions are found in, for example, \cite{Froyland1997, ColbrookTownsend2021, SchuetteKlusHartmann2022}.

The methodology here most closely mirrors that of a combination of \cite{Froyland2013} and \cite{Fernex2021}, where the goal is to construct a discretization of the generator (continuity / Fokker-Planck operator). Furthermore, to make headway on a direct discretization of statistics coming from a partial differential equation, we take on a field-theoretic perspective such as that of \cite{Hopf1952}. 

The rest of the paper is organized as follows: In Section \ref{theory}, we discuss the underlying theory and approximations. The approach is heavily inspired by \cite{Hopf1952, ChaosBook}. The primary idea is to discretize the equations for the statistics (an ``Eulerian" quantity) by using the equation for the dynamics (a ``Lagrangian" quantity). Further approximations are then made to calculate observables of interest.

In Section \ref{methodology}, we introduce a data-driven method with quantified uncertainties for calculating the approximate generator. The method can be applied to arbitrarily long time-series data,  dynamical systems with a large state-space (1,000,000+ degrees of freedom) and further provides uncertainty estimates on the entries of the discretized generator.

Section \ref{applications} goes through two examples: The first serves as an introduction to the concepts and limitations of the method using \cite{Lorenz1963}. The second example uses the compressible Euler equations with gravity and rotation on a thin-shell sphere in the atmospheric benchmark case proposed by \cite{HS_1994}. The system exhibits turbulence and serves as a proxy for Earth's climatology. One sees that even a coarse discretization of statistics captures many features of the original system.

For those simply interested in what the calculations enable, Sections \ref{theory}-\ref{methodology} are safely skipped in favor of Section \ref{applications}. Furthermore, the Appendices \ref{koopman_modes},  \ref{symmetries}, \ref{held_suarez}, expand on the text by discussing Koopman modes, the incorporation of simple symmetries, and the numerical discretization of the compressible Euler equations, respectively.

\section{Theory}
\label{theory}

\subsection{Finite Dimensional Dynamical System}
We start with a generic continuous time dynamical system in $d$-dimensions given by
\begin{align}
\label{dynamics}
\dot{\pmb{s}} = \pmb{U}(\pmb{s}) 
\end{align}
where $\pmb{s}(t): \mathbb{R} \rightarrow \mathbb{R}^d$ is the state of the system and $\pmb{U}: \mathbb{R}^d \rightarrow \mathbb{R}^d$ is the evolution equation. Equation \ref{dynamics} provides a succinct rule for determining the evolution of a dynamical system; however, uncertain initial conditions hemorrhage future predictions in the presence of chaos, \cite{Lorenz1963}. It is, therefore, more natural to study the statistical evolution of probability densities as in \cite{Hopf1952}. Thus we focus not on the $d-$dimensional ordinary differential equation given by Equation \ref{dynamics} but rather the $d-$dimensional partial differential equation that governs the evolution of probability densities in state space.

To do so, we denote fixed vector in state space by $\pmb{\mathscr{s}} \in \mathbb{R}^d$, the components of the state $\pmb{\mathscr{s}}$ by $\pmb{\mathscr{s}} = (\mathscr{s}_1, \mathscr{s}_2, ..., \mathscr{s}_d)$, and the components of the evolution rule by $\pmb{U} = (U_1, U_2, ... U_d)$. To clarify the conceptual differences, we do not commit the usual notational crime of setting $s = \mathscr{s}$. The evolution equation for the statistics of Equation \ref{dynamics}, as characterized by a probability distribution function,
\begin{align}
\mathcal{P} = \mathcal{P}(\mathscr{s}_1, \mathscr{s}_2, ..., \mathscr{s}_d, t) = \mathcal{P}(\pmb{\mathscr{s}}, t),
\end{align}
is given by the continuity equation 
\begin{align}
\label{statistics}
    \partial_t \mathcal{P} + \sum_{i=1}^d \frac{\partial}{\partial \mathscr{s}_i} \left( U_i(\pmb{\mathscr{s}}) \mathcal{P} \right) = 0.
\end{align}
The above equation is a statement of probability conservation. It is precisely analogous to the mass continuity equation from the compressible Navier-Stokes equations. However, the ``mass" is being interpreted as a probability density. The distribution, $\mathcal{P}$, is guided by the flow dynamics $\pmb{U}$ to likely regions of state space. Thus, our focus is not on the ``Lagrangian" view given by dynamics Equation \ref{dynamics} but rather the ``Eulerian" view as given by \ref{statistics}. 

\subsection{Infinite-Dimensional Dynamical System}
When the underlying dynamical system is a partial differential equation, we assume that a suitably well-defined discretization exists to reduce it to a formally $\mathbb{R}^d$ dimensional dynamical system. We contend ourselves to the study of the statistics of the  $\mathbb{R}^d$ approximation. One hopes that different discretizations lead to similar statistical statements of the underlying partial differential equation; thus, it is worth introducing notation for the analogous continuity equation for a partial differential equation, as was done by \cite{Hopf1952}. 

We take care with notation. The $d$-dimensional vector from before now becomes a vector in function space whose components are labeled by a continuous index, $\pmb{x}$, a position in a domain $\Omega$, and discrete index $j$, the index for the field of interest. Thus the component choice $s_i(t)$ for a fixed index $i$ is analogous to $s_{(\pmb{x}, j)}(t)$ for a fixed position $\pmb{x}$ and field index $j \in \{1, ..., d_s \}$,  e.g.  $d_s = 3$ for the three velocity components of the incompressible Navier-Stokes. In the discrete case, the single index $i$ loops over all velocity components and all simulation grid points.   We choose to forego the usual convention of using $\pmb{\tilde{s}}(\pmb{x}, t)$ since, $\pmb{s}: \mathbb{R} \rightarrow \mathcal{X}$ where $\mathcal{X}$ is a function space. The other notation instead suggestions a mapping of the form $\pmb{\tilde{s}}: \Omega \times \mathbb{R} \rightarrow \mathbb{R}^{d_s}$.

Specifically, we consider a partial differential equation for a state $\pmb{s}$ defined over a domain $\Omega$, with suitable boundary conditions,
\begin{align}
\label{functional_dynamics}
    \partial_t \pmb{s} = \pmb{\mathcal{U}}[\pmb{s}]
\end{align}
where the operator $\pmb{\mathcal{U}}: \mathcal{X} \rightarrow \mathcal{X}$ characterizes the evolution of system. The component of $\pmb{\mathcal{U}}$ at position $\pmb{x}$ and field index $j$ is denoted by $\mathcal{U}_{(\pmb{x}, j)}$. 

The analogous evolution for the probability  density functional,
\begin{align}
\mathcal{P} =\mathcal{P}[\mathscr{s}_{(\pmb{x},1)}, \mathscr{s}_{(\pmb{x},2)}, ..., \mathscr{s}_{(\pmb{x},d_s)}, t] = \mathcal{P}[\pmb{\mathscr{s}}, t]   
\end{align}
is denoted by
\begin{align}
\label{functional_statistics}
    \partial_t \mathcal{P} + \sum_{j=1}^{d_s} \int_{\Omega} d \pmb{x} \frac{\delta }{\delta \mathscr{s}_{(\pmb{x},j)} } \left( \mathcal{U}_{(\pmb{x},j)}[\pmb{\mathscr{s}}] \mathcal{P} \right) = 0.
\end{align}
The sum in Equation \ref{statistics} is replaced by both an integral over position indices and a sum over field indices in \ref{functional_statistics}. Furthermore, the partial derivatives are replaced by variational derivatives. The variational derivative is being used in the physicist's sense, that is to say,  
\begin{align}
\frac{\delta \mathscr{s}_{(\pmb{x }, i)} }{\delta \mathscr{s}_{ (\pmb{y}, j)} } = \delta (\pmb{x} - \pmb{y}) \delta_{ij} \Leftrightarrow \frac{\partial \mathscr{s}_{i'}}{\partial \mathscr{s}_{j'}} = \delta_{i'j'}
\end{align}
in analogy to the discrete identity. In the typical physics notation, it is common to drop the dependence on the position $\pmb{x}$ and explicitly write out the field variable in terms of its components (as opposed to the indexing that we do here), e.g.,
\begin{align}
 \frac{\delta }{\delta \mathscr{s}_{ (\pmb{y },1)} } \sum_{i=1}^{d_s} \int_{\Omega} d\pmb{x} (\mathscr{s}_{(\pmb{x}, i)})^2 = 2 \mathscr{s}_{(\pmb{y},1)}  \Rightarrow  \frac{\delta }{\delta u} \int_{\Omega}  (u^2 + v^2 + w^2)= 2 u
\end{align}
for the prognostic variables $u,v,w$ of the incompressible Navier-Stokes equations.

To derive Equation \ref{functional_statistics}, we suppose that Equation \ref{dynamics} is a discretization of Equation \ref{functional_dynamics}. Starting from Equation \ref{statistics}, first introduce a control volume in at index i as $\Delta \pmb{x}_i$ to rewrite the equation as 
\begin{align}
\label{statistics_2}
    \partial_t \mathcal{P} + \sum_{i=1}^d \Delta \pmb{x}_i \frac{1}{\Delta \pmb{x}_i}\frac{\partial}{\partial \mathscr{s}_i} \left( U_i(\pmb{\mathscr{s}}) \mathcal{P} \right) = 0.
\end{align}
In the ``limit", we have 
\begin{align}
\sum_{i=1}^d \Delta \pmb{x}_i \frac{1}{\Delta \pmb{x}_i}\frac{\partial}{\partial \mathscr{s}_i} \rightarrow \sum_{j=1}^{d_s} \int_{\Omega} d \pmb{x} \frac{\delta }{\delta \mathscr{s}_{(\pmb{x},j)} }  \text{ and } U_i \rightarrow \mathcal{U}_{(\pmb{x},j)}
\end{align}

The focus of this work is on methods for discretizing Equations \ref{statistics} and Equation \ref{functional_statistics} on subsets of state space that are typically thought of as chaotic or turbulent given only trajectory information from Equations \ref{dynamics} and Equation \ref{functional_dynamics}, respectively. We use modern data-driven methods of accumulating statistics from trajectory data similar to \cite{Klus2016, Fernex2021}, but use a different method of partitioning that allows us to generalize to the functional setting given by Equation \ref{functional_statistics} and with arbitrary amounts of time series data. 

\subsection{Finite Volume Discretization}
\label{finite_volume}

To focus our discussion, we use the finite-dimensional setting. However, the arguments apply mutatis mutandi to the infinite-dimensional one\footnote{This is, of course, assuming that there a sense in which limits are justified. Furthermore, infinitesimal volumes in state space in the discrete setting $d \pmb{\mathscr{s}}$ are instead denoted by $\mathcal{D}[\pmb{\mathscr{s}}]$ in the function space setting.}. We first assume that the underlying dynamics are on a chaotic attractor associated with a compact subset of state space $\mathcal{M} \subset \mathbb{R}^d$. We introduce $N$ partitions of $\mathcal{M}$ which we denote by $\mathcal{M}_n$ for $n = 1, ..., N$. 

The coarse-grained discretization variables $P_n$ are
\begin{align}
\int_{\mathcal{M}_n}  d \pmb{\mathscr{s}}  \mathcal{P} =  \int_{\mathcal{M}_n}  \mathcal{P} = P_n
\end{align}
as is common in finite volume methods. We will drop the infinitesimal state space volume $d \pmb{\mathscr{s}} $ when unambiguous. Here $P_n(t)$ is the probability in time of being found in the subset of state space $\mathcal{M}_n$ at time $t$. Integrating Equation \ref{statistics} with respect to the partitions yields 
\begin{align}
\label{coarse_graining}
 \frac{d}{d t} P_n =    \int_{\mathcal{M}_n} \left[ \sum_{i=1}^d \frac{\partial}{\partial \mathscr{s}_i} \left( U_i(\pmb{\mathscr{s}}) \mathcal{P} \right)  \right] = \int_{\partial \mathcal{M}_n} \pmb{U} \cdot \hat{\pmb{n}} \mathcal{P}
\end{align}
where $\partial \mathcal{M}_n$ is the boundary of the partition and $\hat{n}$ is a normal vector. The art of finite volume methods comes from expressing the right-hand side Equation \ref{coarse_graining} in terms of the coarse-grained variables $P_n$ through a suitable choice of numerical flux. 

We go about calculating the numerical flux in a roundabout way. We list some desiderata for a numerical discretization
\begin{enumerate}
    \item The discrete equation is expressed in terms of the instantaneous coarse-grained variables $P_n$. 
    \item The discrete equation is linear, in analogy to the infinite-dimensional one. 
    \item The equation must conserve probability. 
    \item Probability must be positive at all times.
\end{enumerate}
The first two requirements state,
\begin{align}
\int_{\partial \mathcal{M}_n} \pmb{U} \cdot \hat{\pmb{n}} \mathcal{P} \approx \sum_{m} \mathcal{Q}_{nm} P_m
\end{align}
for some matrix $\mathcal{Q}$. Thus we want an equation of the form 
\begin{align}
    \frac{d}{dt} \hat{P}_n &= \sum_m \mathcal{Q}_{nm} \hat{P}_m
\end{align}
We introduced a ``hat" to distinguish the numerical approximation, $\hat{P}_n$, with the exact solution $P_n$.
The third requirement states that 
\begin{align}
    \sum_n \hat{P}_n = 1 \Rightarrow \frac{d}{d t} \sum_n \hat{P}_n = 0 = \sum_{mn} \mathcal{Q}_{nm} \hat{P}_m
\end{align}
for each $\hat{P}_m$, thus 
\begin{align}
    \sum_{n} \mathcal{Q}_{nm} = 0
\end{align}
for each m, i.e., the columns of the matrix must add to zero. Moreover, the last requirement states that the off-diagonal terms of $Q_{nm}$ must all be positive. To see this last point, we do a proof by contradiction. Suppose there is a negative off-diagonal entry, without loss of generality, component $\mathcal{Q}_{21}$. Then if at time zero our probability vector starts at $\hat{P}_1(0) = 1$ and $\hat{P}_n(0) = 0$ for $n > 1$, an infinitesimal timestep $dt$ later we have
\begin{align}
\hat{P}_2(dt ) =  dt \sum_{m} Q_{2m} \hat{P}_m(0) = Q_{21} < 0,
\end{align}
a contradiction since probabilities must remain positive at all times. Of all the requirements, it is this fourth one that is most readily abandoned since it is not possible to have a higher-order discretization that is both positivity preserving and linear \cite{Shu_2011}. 

These four requirements, taken together, are enough to identify the matrix $\mathcal{Q}$ as the generator of a continuous-time Markov process with finite state space. This observation forms the backbone of the data-driven approach towards discretizing equation \ref{statistics} and \ref{functional_statistics}. The diagonal entries of the matrix are related to the average amount of time spent in a partition, and the off-diagonal entries within a column are proportional to the probabilities of entering a given partition upon exit of a partition. The implication is that we construct the numerical fluxes on the boundary through Monte Carlo integration of the equations of motion. 

Intuitively, as a dynamical system enters through a partition of state space, it becomes associated with being within that partition. The time a trajectory spends within a partition is called the holding time from whence it will eventually exit to some other partition in state space. A sufficiently long integration of the equations of motion constructs the holding time distributions and exit probabilities to different partitions in state space. Furthermore, to perform calculations, we associate each region of state space with a ``cell-center" which, in this paper, we call a Markov state. The Markov state will serve as a center of a delta function approximation to the distribution in that region of state space. With the transition matrix $\mathcal{Q}$ and the Markov states associated with a partition, we can perform calculations of moments, steady-state distributions, and autocorrelations of any variable of interest.

\subsection{Time versus Ensemble Calculations}
\label{ensemble_vs_time}
At this point, we have discussed the equation for statistics of a dynamical system, the notation for the infinite-dimensional case, and how to associate a continuous time Markov process with a finite volume discretization of the continuity equation. We now discuss how to perform statistical calculations from the discretization and how we will confirm that the discretization captures statistics of the underlying continuity equation. In short, we compare temporal averages to ensemble averages and analogous calculations for autocorrelations. 

We must introduce additional notation. As stated before, we assume that an ergodic chaotic attractor exists so that there is a unique invariant measure, which we denote by $\mathscr{P}(\pmb{\mathscr{s}})$. The conditional invariant measure with respect to a partition $\mathcal{M}_n$ is denoted by $\mathscr{P}(\pmb{\mathscr{s}} | \mathcal{M}_n)$ and the probability of a state being found in a partition $\mathcal{M}_n$ is $\mathbb{P}(\mathcal{M}_n)$ so that the invariant measure is decomposed as
\begin{align}
\label{conditional_decomposition}
\mathscr{P}(\pmb{\mathscr{s}}) &= \sum_n \mathscr{P}(\pmb{\mathscr{s}} | \mathcal{M}_n) \mathbb{P}(\mathcal{M}_n).
\end{align}
 In addition, we introduce notation for the transfer operator $\mathscr{T}^\tau$, which defined through the relation
\begin{align}
    \mathcal{P}(\pmb{\mathscr{s}}, t + \tau ) = \mathscr{T}^\tau \mathcal{P}(\pmb{\mathscr{s}}, t ) 
\end{align}
where $\mathcal{P}$ is a solution to Equation \ref{statistics}. Thus the transfer operator is an instruction to evolve the density, $\mathcal{P}$, via Equation \ref{statistics} to a time $\tau$ in the future. Furthermore,
\begin{align}
    \lim_{\tau \rightarrow \infty} \mathscr{T}^\tau \mathcal{P}(\pmb{\mathscr{s}}, t ) = \mathscr{P}( \pmb{\mathscr{s}})
\end{align}
for arbitrary densities $\mathcal{P}(\pmb{\mathscr{s}}, t )$, including $\delta ( \pmb{\mathscr{s}} - \pmb{\mathscr{s}}')$ for an initial state $\pmb{\mathscr{s}}' \in \mathcal{M}$ from our assumption of ergodicity\footnote{We are being sloppy with limits here, but this should be understood where the limit to a delta function density is the last limit taken. }. 

For an observable $g: \mathcal{M} \rightarrow \mathbb{R}$, we calculate long time averages
\begin{align}
\langle g \rangle_T =   \lim_{T \rightarrow \infty} \frac{1}{T} \int_0^T g( \pmb{s} (t)) dt
\end{align}
and compare to ensemble averages
\begin{align}
\langle g \rangle_E = \int_{\mathcal{M}}  g(\pmb{\mathscr{s}} ) \mathscr{P}(\pmb{\mathscr{s}})
\end{align}

Furthermore, we compare time-correlated observables. The time series calculation is
\begin{align}
R_T(g, \tau) &= \lim_{T \rightarrow \infty} \frac{1}{T} \int_0^T g(\pmb{s}(t+\tau))g(\pmb{s}(t)) dt,
\end{align}
from whence we obtain the autocovariance, $C_T$, and autocorrelation, $\tilde{C}_T$,
\begin{align}
C_T(g, \tau) \equiv R_T(g, \tau) - \langle g \rangle_T^2  
\text{ and } 
\tilde{C}_T(g, \tau) \equiv C_T(g, \tau) / C_T(g, 0).
\end{align}

The ensemble average version takes more explanation. We correlate a variable $g(s(t))$ with $g(s(t+\tau))$, which involves the joint distribution of two variables. To state the ensemble average version, we first review a fact about random variables $X, Y$ with joint density $\rho(x,y)$, conditional density $\rho(x |y)$, and marginal density $\rho_y(y)$. The expected value, the correlation of two observables, is calculated as
\begin{align}
    \langle g(X) g(Y) \rangle &= \iint dx dy g(x) g(y) \rho(x,y)
    = \iint dx dy g(x) g(y) \rho(x|y)\rho_y(y) 
    \\
    &= \int dy g(y) \rho_y(y) \left[ \int dx g(x) \rho(x|y) \right].
\end{align}
To translate the above calculation to the present case, we consider $\rho_y$ as the invariant measure, $\mathscr{P}$. The conditional distribution $\rho(x|y)$ is thought of as the probability density at a time $\tau$ in the future, given that we know that it is initially at state $\pmb{\mathscr{s}}$ at $\tau = 0$.

Thus in our present case, $\rho(x|y)$ becomes  $\mathscr{T}^\tau \delta ( \pmb{\mathscr{s}} - \pmb{\mathscr{s}}')$  where the $\delta$ function density is a statement of the exact knowledge of the state at time $\tau = 0$. In total, the ensemble time-autocorrelation is calculated as
\begin{align}
    R_E(g, \tau) 
    &= \int_{\mathcal{M}} d\pmb{\mathscr{s}}' g(\pmb{\mathscr{s}}' )   \mathscr{P}( \pmb{\mathscr{s}}' )\left[ \int_{\mathcal{M}} d\pmb{\mathscr{s}} g(\pmb{\mathscr{s}}) \mathscr{T}^\tau \delta ( \pmb{\mathscr{s}} - \pmb{\mathscr{s}}') \right].
\end{align}
The autocovariance, $C_E$, and autocorrelation, $\tilde{C}_E$, are
\begin{align}
C_E(g, \tau) \equiv R_E(g, \tau) - \langle g \rangle_E^2  
\text{ and } 
\tilde{C}_E(g, \tau) \equiv C_E(g, \tau) / C_E(g, 0).
\end{align}

These calculations summarize the exact relations we wish to compare. However, first, we will approximate the temporal averages via long-time finite trajectors and the ensemble averages via the finite-volume discretization from section \ref{finite_volume}. 

\subsection{Approximations to Time versus Ensemble Calculations}
The prior section represents the mathematical ideal with which we would like to perform calculations; however, given that we use a data-driven construction, we are faced with performing calculations in finite-dimensional spaces and over finite-dimensional time.

Given the time series of a state at evenly spaced times at times $t_n$ for $n = 1$ to $N_t$ with time spacing $\Delta t$, we approximate the mean and long time averages of an observable $g$ as 
\begin{align}
\langle g \rangle_T &\approx \frac{1}{N_t} \sum_{n = 1}^{N_t} g(\pmb{s}(t_n)) 
\\
\label{time_autocorrelation_approximation}
R_T(g, \tau) 
&\approx  \frac{1}{N_t'} \sum_{n = 1}^{N_t'} g(\pmb{s}(t_n + \text{round}(\tau / \Delta t) \Delta t)) g(\pmb{s}(t_n)) 
\end{align}
where the $\text{round}$ function computes the closest integer and $N_t' = N_t - \text{round}(\tau / \Delta t)$.

We use the construction in section \ref{finite_volume} to calculate ensemble averages. Recall that in the end, we had approximated the generator of the process with a matrix $Q$, which described the evolution of probabilities associated with partitions of state space. In addition, we select a state $\pmb{\sigma}^n$ associated with a partition $\mathcal{M}_n$ as the ``cell-center" in order to perform calculations. We use superscripts to denote different states since subscripts are reserved for the evaluation of the component of a state. Furthermore, we do not require the Markov state $\pmb{\sigma}^n$ to be a member of the partition $\mathcal{M}_n$. For example, we could choose the $\pmb{\sigma}^n$ as fixed points of the dynamical system or a few points along a periodic orbit within the chaotic attractor $\mathcal{M}$.

The ensemble average of an observable is calculated by making use of the decomposition of the invariant measure Equation \ref{conditional_decomposition}, but then approximating
\begin{align}
    \mathscr{P}(\pmb{\mathscr{s}} | \mathcal{M}_n) \approx \delta(\pmb{\mathscr{s}} - \pmb{\sigma}^n )
\end{align}
which is a simple but crude approximation. Thus the ensemble averages are calculated as 
\begin{align}
    \langle g \rangle_E &= \int_{\mathcal{M}} g( \pmb{\mathscr{s}} ) \left[ \sum_n \mathcal{P} \left(\pmb{\mathscr{s}} | \mathcal{M}_n \right) \mathbb{P}(\mathcal{M}_n) \right]  
    =  \sum_n  \left[ \int_{\mathcal{M}} g( \pmb{\mathscr{s}} ) \mathcal{P} \left(\pmb{\mathscr{s}}  | \mathcal{M}_n \right)  \right] \mathbb{P}(\mathcal{M}_n) 
    \\
    \label{ensemble_average_approximation}
    &\approx \sum_n  \left[ \int_{\mathcal{M}} g( \pmb{\mathscr{s}} ) \delta \left( \pmb{\mathscr{s}} - \pmb{\sigma}^n \right)\right] \mathbb{P}(\mathcal{M}_n) = \sum_n g( \pmb{\sigma}^n ) \mathbb{P}(\mathcal{M}_n) 
\end{align}

For the ensemble average version of time auto-correlations, we must, in addition to approximating the invariant measure, approximate the transfer operator acting delta function density of the state, $\mathscr{T}^\tau \delta(\pmb{\mathscr{s}} - \pmb{\mathscr{s}}')$.
We calculate
\begin{align}
    R_E(g, \tau) 
    &= \int_{\mathcal{M} } \int_{\mathcal{M} } d\pmb{\mathscr{s}}'  d\pmb{\mathscr{s}} g(\pmb{\mathscr{s}}' )   \mathscr{P}( \pmb{\mathscr{s}}' ) g(\pmb{\mathscr{s}}) \mathscr{T}^\tau \delta ( \pmb{\mathscr{s}} - \pmb{\mathscr{s}}')
    \\
    &\approx \sum_{n} \int_{\mathcal{M} } \int_{\mathcal{M} } d\pmb{\mathscr{s}}'  d\pmb{\mathscr{s}} g(\pmb{\mathscr{s}}' )   \delta(\pmb{\mathscr{s}}' - \pmb{\sigma}^n ) \mathbb{P}(\mathcal{M}_n) g(\pmb{\mathscr{s}}) \mathscr{T}^\tau \delta ( \pmb{\mathscr{s}} - \pmb{\mathscr{s}}')
    \\
    &= \sum_{n} g(\pmb{\sigma}^n ) \mathbb{P}(\mathcal{M}_n) \int_{\mathcal{M} }  d\pmb{\mathscr{s}} g(\pmb{\mathscr{s}}) \mathscr{T}^\tau \delta ( \pmb{\mathscr{s}} - \pmb{\sigma}^n)
\end{align}
then additionally approximate 
\begin{align}
\mathscr{T}^\tau \delta(\pmb{\mathscr{s}}- \pmb{\sigma}^n) 
&\approx \sum_{m=1}^{N} \delta(\pmb{\sigma}^m - \pmb{\mathscr{s}})  [\exp(Q \tau)]_{mn} .
\end{align}
The matrix exponential $\exp(Q \tau)$ is the analogous Perron-Frobenius/Transfer operator for the discrete system. The matrix is a (column) stochastic matrix whose entries sum to one. The intuition behind the approximation is to treat the forward evolution of the transfer operator for delta distribution centered at state $\pmb{\sigma}^n$ as a weighted sum of delta functions centered at state $\pmb{\sigma}^m$. Putting together the pieces results in 
\begin{align}
\label{ensemble_autocorrelation_approximation}
    R_E(g, \tau) 
    &\approx  
    \sum_{n = 1}^{N} g(\pmb{\sigma}^n) \mathbb{P}(\mathcal{M}_n) \left[ \sum_{m=1}^{N}  g( \pmb{\sigma}^m)  [\exp(Q \tau)]_{mn}  \right]
\end{align}

In addition, all covariances and correlations are calculated by making use of the above approximations. This review completes the discussion of how to approximate ensemble averages and covariances from the finite volume discretization of the generator. However, it remains to be shown how to construct the matrix and Markov states from data. The construction of the generator is the subject of the following section.

\section{Methodology}
\label{methodology}
In this section, we outline the general approach to constructing the approximate generator in terms of trajectory data. The most critical component of a discretization comes from defining an embedding function $\mathcal{E}: \mathbb{R}^d  \rightarrow \{1, 2, ..., N \}$ which maps an arbitrary state $\pmb{\mathscr{s}} \in \mathbb{R}^d$ to an integer $n \in  \{1, 2, ..., N \}$. This function implicitly defines a partition through the relation 
\begin{align}
\mathcal{M}_j = \{ \pmb{\mathscr{s}} : \mathcal{E}(\pmb{\mathscr{s}}) = j \text{ for  each } \pmb{\mathscr{s}} \in \mathbb{R}^d \} \cap \mathcal{M}.
\end{align}
The intersection with the manifold $\mathcal{M}$ is critical to the success of the methodology.

Furthermore, the Markov states (cell-centers) are chosen to satisfy $\mathcal{E}(\pmb{\sigma}^n) = n$ for each $n \in \{1, ..., N \}$. There is an extraordinary amount of freedom in defining the embedding function, and we will go through three examples in Section \ref{applications}. We also comment on practical considerations and generalizations in Section \ref{discussion}. One can simultaneously solve for an embedding function and Markov states using a K-means algorithm, see \cite{Lloyd1982}, but we do not wish to restrict ourselves to that choice here. The embedding function is a classifier (in the machine learning sense) for different flow states with integers at the category labels. For now, we will assume that such a function is given and focus on constructing the generator $Q$. 

The Markov embedding function $\mathcal{E}$ transforms
dynamical trajectories into sequences of integers which we interpret realization of a Markov process with finite state space. At this stage, traditional methods can be employed to construct a Transfer/Perron-Frobenius operator from data, see \cite{Klus2016, Fernex2021}. Given that our interest is in constructing a continuous time Markov process, the algorithm will be a modification in line with \cite{Froyland2013}. To construct $Q$, two quantities must be calculated for each partition
\begin{enumerate}
    \item The holding times: The amount of time a dynamical trajectory stays in partition $\mathcal{M}_n$ before exiting.
    \item The exit probabilities: The probability of moving from partition $\mathcal{M}_j$ to $\mathcal{M}_i$ upon exiting the partition $\mathcal{M}_j$.
\end{enumerate}
Let $T_j$ be the distribution of holding times associated with partition $j$ and $E_{ij}$ denote entries of the exit probability matrix. By our convention $\sum_j E_{ij} = 1$ and $E_{ii} = 0$ for all $i$. The entries of the matrix $Q_{ij}$ are constructed as follows 
\begin{align}
Q_{ij} = E_{ij} / \langle T_j \rangle \text{ for } i \neq j \text{ and } Q_{jj} = - 1/ \langle T_j \rangle \text{ for } j \in \{1, ..., N \}
\end{align}
where $\langle T_j \rangle$ denotes the expected value of the holding time distribution of partition $j$. 

In the subsections, we outline an empirical construction of the matrix from finite data and a Bayesian approach that incorporates uncertainty due to finite sampling effects. With the latter approach, we do not treat the entries of the $Q_{ij}$ matrix as deterministic numbers but rather as distributions. The result is a random matrix representation of the generator that incorporates uncertainty. 

\subsection{Empirical Construction}
\label{empirical_construction}
We start with an empirical construction of the generator. It suffices focus on partition $j$ associated with the $j$'th column of the matrix $Q_{ij}$. To calculate the empirical holding time distribution and empirical mean, we count up how often we see state $j$ before transitioning to state $i \neq j$. For example, suppose that we have three states, $j=1$, and consider the following sequence of integers given by a Markov embedding applied to a time series with $\Delta t$ spacing in time,
\begin{align}
1,  1, 1, 2, 2, 1, 1, 3, 1, 2, 1, 1.
\end{align}
We group the sequence as follows
\begin{align}
(1,  1, 1), 2, 2, (1, 1), 3, (1), 2, (1, 1)
\end{align}
to determine the holding times. Thus, the holding times for state 1 would be 
\begin{align}
    3 \Delta t, 2\Delta t, \Delta t, 2\Delta t
\end{align}
whose empirical expected value is $2 \Delta t $ implying a transition rate $1 / (2 \Delta t)$.

To calculate exit probabilities for partition $j$, we count how often we see transitions to partitions $i$ and divide by the total number of transitions. In the example, to calculate the exit probabilities for partition $1$ into partition $2$ or $3$, we group them together as follows
\begin{align}
1,  1, (1, 2,) 2, 1, (1, 3), (1, 2), 1, 1
\end{align}
Thus we saw three exits, two of which went to state $2$ and one of which went to state $3$; hence the exit probabilities are $E_{21} = 2/3$ and $E_{31} = 1/3$. 

The rest of the states are constructed analogously to produce the matrix 
\begin{align}
Q = \frac{1}{\Delta t}\begin{bmatrix}
-1/2         &   \text{ }2/3  &  \text{ } 1 \\
\text{ }1/3  & -2/3           &    0 \\
\text{ }1/6  &  0             & -1
\end{bmatrix}
\end{align}

As currently implemented, the generator is only accurate to order $\Delta t$ since we do not interpolate in time to find the ``exact" holding time. We do not preoccupy ourselves with improving this since we believe that the primary source of error comes from finite sampling effects. In the following section, we augment the empirical construction with uncertainty estimates based on finite sampling and a Bayesian framework.

\subsection{Bayesian Construction}
\label{bayesian_construction}
We need four ingredients to enable a Bayesian construction of the generator. 
\newline 
\begin{enumerate}
    \item A likelihood distribution for the holding times
    \item A prior distribution for the transition rates associated with the holding times
    \item A likelihood distribution for the exit probabilities
    \item A prior distribution for the probability values associated with the exit probabilities
\end{enumerate}
$ $
\newline
We make assumptions compatible with drawing from a continuous time Markov process for each column independently
\newline 
\begin{enumerate}
    \item The likelihood distribution for the holding times is exponentially distributed with rate parameter $\lambda_i$
    \item The likelihood distribution for exit probabilities is a Multinomial Distribution with parameters $\vec{p} \in [0, 1]^{N-1}$ satisfying the relation $\sum_{i=1}^{N-1} p_i = 1$
    \item The prior distribution for the rate parameter of the exponential distribution is distributed according to the gamma distribution with parameters $(\alpha, \beta)$, denoted by $\Gamma(\alpha, \beta)$
    \item The prior distribution for the probabilities in the Multinomial distribution comes from a Dirichlet distribution with parameter vector $\vec{\alpha}$ of length $N-1$, which we denote by Dirichlet($\vec{\alpha}$)
\end{enumerate}
$ $
\newline 
The distributions are conjugate priors which allows for the posterior distribution to come from the same family as the prior; see \cite{BayesianDataAnalysis}. For example, under this construction, a $3 \times 3$ matrix will always be of the form 
\begin{align}
Q = \begin{bmatrix}
- 1              &  [\vec{D}_{2}]_1    &   [\vec{D}_{3}]_1  \\
[\vec{D}_{1}]_1  & -1                  &   [\vec{D}_{3}]_2  \\
[\vec{D}_{1}]_2  &  [\vec{D}_{2}]_2     &  - 1
\end{bmatrix}
\begin{bmatrix}
G_1   &  0  &    0 \\
0  &  G_2      &    0 \\
0  &   0    &  G_3
\end{bmatrix}
\end{align}
where $G_i \sim \Gamma(\alpha_i, \beta_i)$, $\vec{D}_{i} \sim \text{Dirichlet}(\vec{\alpha}_i)$, and $[\vec{D}_{i}]_j$ denotes the $j$'th component of the random vector $\vec{D}_{i}$.

The parameters $(\alpha_i, \beta_i)$ and $\vec{\alpha}_i$ are updated according to Bayes rule for each column upon data acquisition. For example, suppose that we have observed the following empirical counts associated with partition $i$
\begin{enumerate}
\item $M$ exits from partition $i$
\item $[\vec{M}]_j$ exits from partition $i$ to partition $j$.
\item $\hat{T}_1, \hat{T}_2, ..., \hat{T}_M$ empirically observed holding times
\end{enumerate}
and that we start with $(\alpha^0, \beta^0)$ and $\vec{\alpha}^0$ as the parameters for our prior distribution. The relation $\sum_j [\vec{M}]_j = M$ holds\footnote{Technically, there can be an ``off-by-one" error here which we ignore for presentation purposes.}. The posterior distribution parameters $(\alpha^1, \beta^1)$ and $\vec{\alpha}^1$ are 
\begin{align}
\alpha^1 &= \alpha^0 + M \text{ , }
\beta^1  = \beta^0 + \sum_{i}^M \hat{T}_i \text{, and }
\vec{\alpha}^1 = \vec{\alpha}^0 + \vec{M} .
\end{align}
In the limit that $\alpha^0 , \beta^0$, and $|\vec{\alpha}^0|$ go to zero, then the empirical approach from the prior section agrees with the expected value from the Bayesian approach. 

The current approach is one of many approaches to constructing matrices with quantified uncertainties. However, it is not the only one\footnote{For example, one could account for correlations between columns or treat the likelihood for each exit probability individually as a Bernoulli Distribution with Beta Distribution conjugate prior.}. The current construction is imperfect in many regards (e.g., when holding times do not follow an exponential distribution or the system is not Markovian over infinitesimal steps), but we hold the position that \textit{some} quantification of uncertainty is better than none. We use uncertainty quantification to dismiss spurious results rather than increase confidence in the correctness of an inference. The Perron-Frobenius/Transfer operator can also use the approach here by using the Multinomial Distribution as the likelihood function and using Dirichlet distribution priors.

Examples of using the theory and methodology to construct data-driven approximations to the generator with quantified uncertainties follow.

\section{Applications}
\label{applications}
We apply the methodology from the previous section to two different systems. The dynamics of the first system is the Lorenz equations:
\begin{align}
\label{lorenz_dynamics}
\dot{x} &= - \sigma ( x - y) \\ 
\dot{y} &= - y + (r-z) x \\ 
\dot{z} &= - b z + xy
\end{align}
where we identify $x = s_1$, $y = s_2$, $z = s_3$.  The corresponding continuity equation is given by
\begin{align}
\label{lorenz_statistics}
\partial_t \mathcal{P}
+ \partial_{\mathscr{x}} \left( \left[ - \sigma ( \mathscr{x} - \mathscr{y})\right] \mathcal{P} \right) 
+ \partial_{\mathscr{y}} \left( \left[ - \mathscr{y} + (r-\mathscr{z}) \mathscr{x} \ \right] \mathcal{P} \right) 
+ \partial_{\mathscr{z}}\left( \left[ - b \mathscr{z} + \mathscr{x}\mathscr{y} \right] \mathcal{P} \right)
&= 0
\end{align}
where we use the notation $\mathscr{x}$ for $\mathscr{s}_1$, $\mathscr{y}$ for $\mathscr{s}_2$, and $\mathscr{z}$ for  $\mathscr{s}_3$.

The second system we consider is a Flux-Differencing Discontinuous Galerkin discretization of the compressible Euler equations on the sphere with rotation and gravity. The prognostic variables of choice are density $\rho$, momentum $\rho \pmb{u}$, and total energy $\rho e$. The dynamics are given by the following equations:
\begin{align}
\partial_t \rho &= - \nabla \cdot \left( \rho \pmb{u} \right) 
\\
\partial_t (\rho \pmb{u} ) &= - \nabla \cdot \left( \pmb{u} \otimes  \rho \pmb{u} + p \mathbb{I} \right) -\rho \nabla \Phi + \mathbf{S}_{\rho\pmb{u}}\left(\rho, \rho \pmb{u}, \rho e \right) 
\\
\partial_t (\rho e ) &= - \nabla \cdot \left( \pmb{u} \left( p + \rho e \right) \right) + S_{\rho e}\left(\rho, \rho \pmb{u}, \rho e \right). 
\end{align}
where the details of the source terms, $\mathbf{S}_{\rho\pmb{u}}$ and $S_{\rho e}$, are given in Appendix \ref{held_suarez}, $\Phi$ is the geopotential, and $p$ is pressure. We make the following identifications with prior notation $s_{(\pmb{x}, 1)} = \rho$, $s_{(\pmb{x}, 2)} = \rho u$, $s_{(\pmb{x}, 3)} = \rho v$, $s_{(\pmb{x}, 4)} = \rho w$, $s_{(\pmb{x}, 5)} = \rho e$. The corresponding continuity equation is
\begin{align}
    \partial_t \mathcal{P} &+ \int_{\Omega}  \frac{\delta}{\delta \rho} \left[ - \text{div}\left( \varrho \pmb{\mathscr{u}} \right) \mathcal{P} \right]
    \\
    &+ \int_{\Omega} \frac{\delta}{ \delta \varrho \mathscr{u} } \left[ \left(- \text{div}\left( \pmb{\mathscr{u}}  \varrho \mathscr{u} + p \hat{x} \right) - \varrho \hat{x} \cdot \text{grad}( \Phi ) + S_{\rho u}  \right) \mathcal{P} \right]
    \\
    &+ \int_{\Omega} \frac{\delta}{ \delta \varrho \mathscr{v} } \left[ \left(- \text{div} \left( \pmb{\mathscr{u}}   \varrho \mathscr{v} + p \hat{y} \right) - \varrho \hat{y} \cdot \text{grad}( \Phi )  + S_{\rho v}  \right) \mathcal{P} \right]
    \\
    &+ \int_{\Omega} \frac{\delta}{ \delta \varrho \mathscr{w} } \left[ \left(- \text{div} \left( \pmb{\mathscr{u}}  \varrho  \mathscr{w} + p \hat{z} \right)  -\varrho \hat{z} \cdot \text{grad}( \Phi ) + S_{\rho w}  \right) \mathcal{P} \right]
    \\
    &+ \int_{\Omega} \frac{\delta}{ \delta \varrho \mathscr{e} } \left[ \left(- \text{div} \left( \pmb{\mathscr{u}} \left( p + \varrho \mathscr{e} \right) \right) + S_{\rho e} \right)  \mathcal{P} \right] = 0,
\end{align}
where we make the correspondence
$\mathscr{s}_{(\pmb{x}, 1)} = \varrho$, $\mathscr{s}_{(\pmb{x}, 2)} = \varrho \mathscr{u}$, $\mathscr{s}_{(\pmb{x}, 2)} = \varrho \mathscr{v}$, $\mathscr{s}_{(\pmb{x}, 2)} = \varrho \mathscr{w}$, $\mathscr{s}_{(\pmb{x}, 5)} = \varrho \mathscr{e}$, and $\pmb{\mathscr{u}} = (\mathscr{u}, \mathscr{v}, \mathscr{w})$. Note that the source term $\mathbf{S}_{\rho\pmb{u}}$ has been broken up into three terms $S_{\rho u}$, $S_{\rho v}$, and $S_{\rho w}$. Furthermore we use the ``grad" and ``div" notations to emphasize the connection of mapping functions to functions.

The numerical discretization of the compressible Euler equations is outlined in Appendix \ref{held_suarez}, but is irrelevant for the present purposes. Instead, we consider the system as a finite but high-dimensional space, with $d=1,481,760$ in our specific case.

We choose the Held-Suarez test case for our analysis because it exhibits turbulence, has been extensively studied by the atmospheric community, and is a geophysically relevant configuration that produces wind and temperature patterns similar to those observed on Earth. Moreover, its statistics are robust across multiple discretization strategies, dissipation mechanisms, and equation formulations. It does not exhibit meta-stable states and thus serves as a stringent test on the methodology.

To discretize the first continuity equation, we use three partitions, while we use two partitioning strategies for the latter: first, 400 partitions, and later, 100 partitions. The choice of embedding functions, $\mathcal{E}$, and Markov states $\pmb{\sigma}^n$ is described in the relevant sections.

\subsection{Lorenz: Fixed Point Partitions} 
\label{lorenz_fixed_point_partition}

We choose the classic parameter values $r=28$, $\sigma = 10$, and $b= 8/3$ for the Lorenz system, which is known to exhibit chaotic solutions. Construction of the generator is automated through the methodology of Section \ref{methodology} upon choosing the Markov states $\pmb{\sigma}^n$ and an embedding function $\mathcal{E}$. We use the following fiction to guide our choices:    
 
 It is said that the coherent structures of the flow organize and guide the dynamics of chaos. As a trajectory wanders through state space, it spends a disproportionate time near coherent structures and inherits their properties. The coherent structures then imprint their behavior on the chaotic trajectory, manifesting in ensemble averages. Thus chaotic averages are understood in terms of transitions between simpler structures. This picturesque story motivates the use of fixed points as Markov states,
\begin{align}
\label{fixed_point_1}
\pmb{\sigma}^1  &= [-\sqrt{72} ,-\sqrt{72},  27],
\\
\label{fixed_point_2}
\pmb{\sigma}^2  &= [0, 0, 0],
\\
\label{fixed_point_3}
\pmb{\sigma}^3 &=[\sqrt{72} ,\sqrt{72},  27],
\end{align}
and partitioning state space according to the closest fixed point, 
\begin{align}
\mathcal{E}(\pmb{\mathscr{s}}) &= 
\begin{cases}
1 & \text{ if } \| \pmb{\mathscr{s}} - \pmb{\sigma}^1 \| < \| \pmb{\mathscr{s}} - \pmb{\sigma}^2 \| \text{ and } \| \pmb{\mathscr{s}} - \pmb{\sigma}^3 \| 
\\ 
2 & \text{ if } \| \pmb{\mathscr{s}} - \pmb{\sigma}^2 \| < \| \pmb{\mathscr{s}} - \pmb{\sigma}^1 \| \text{ and } \| \pmb{\mathscr{s}} - \pmb{\sigma}^3 \| 
\\
3 & \text{ if } \| \pmb{\mathscr{s}} - \pmb{\sigma}^3 \| < \| \pmb{\mathscr{s}} - \pmb{\sigma}^1 \| \text{ and } \| \pmb{\mathscr{s}} - \pmb{\sigma}^2 \| 
\end{cases}
\end{align}
where $\| \cdot \|$ denotes the standard Euclidean norm. The embedding function determines the partition by associating a trajectory with the closest fixed point. Stated differently, this partitioning strategy is the intersection of the chaotic attractor, $\mathcal{M}$, with a Voronoi tesselation over the full state space, $\mathbb{R}^3$. We show the partition induced by this choice in Figure \ref{fig:lorenz_partition} from several angles. The regions are color-coded according to the closest fixed points.

\iffigure 
\begin{figure}
\begin{center}
\includegraphics[width=1.0\textwidth]{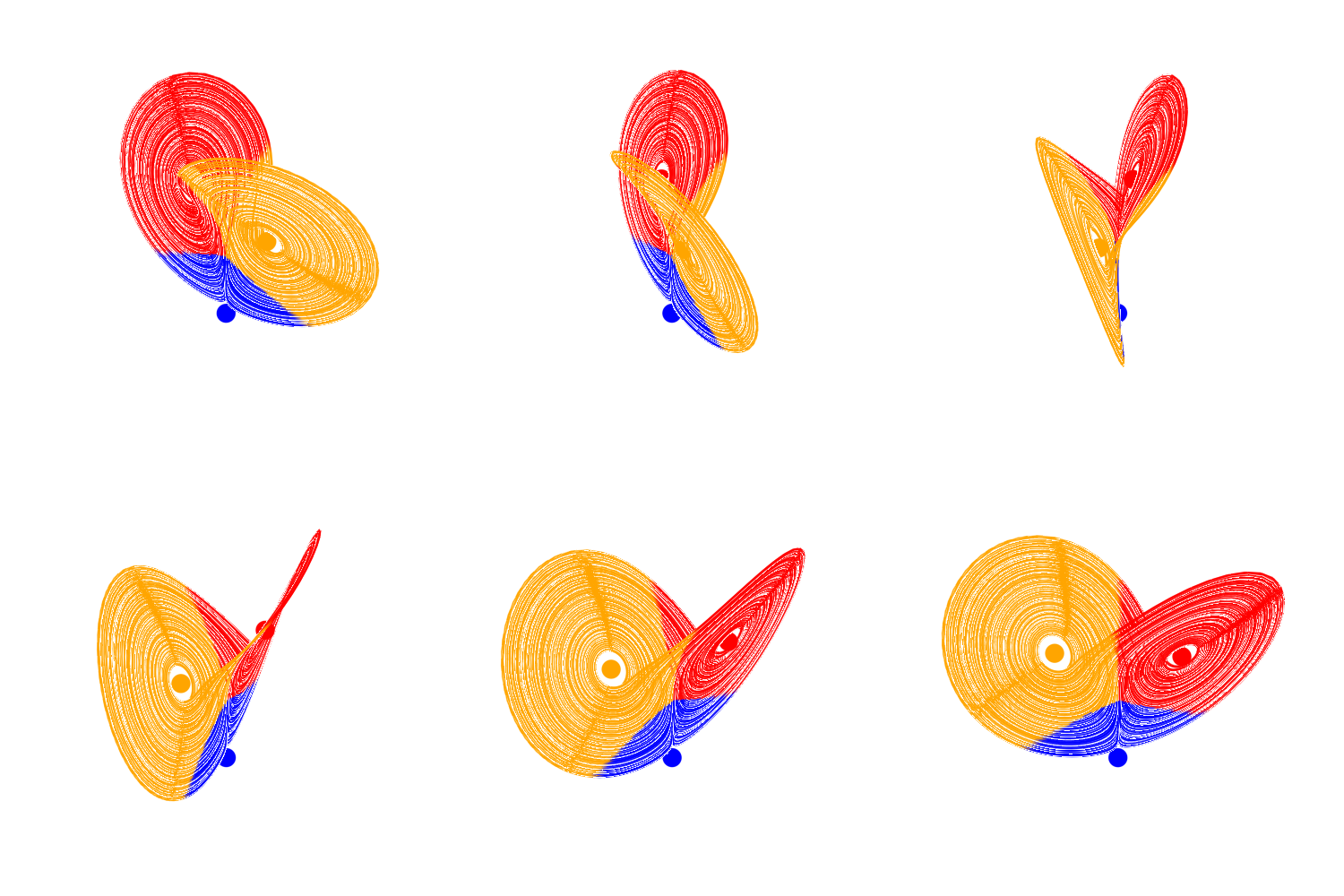}
\caption{Lorenz Fixed Point Partition. Here we show the emerging partition from several angles. The colors correspond to the different partitions associated with trajectories that are ``closest" to a given fixed point.}
\label{fig:lorenz_partition}
\end{center}
\end{figure}
\fi

\iffigure 
\begin{figure}
\begin{center}
\includegraphics[width=1.0\textwidth]{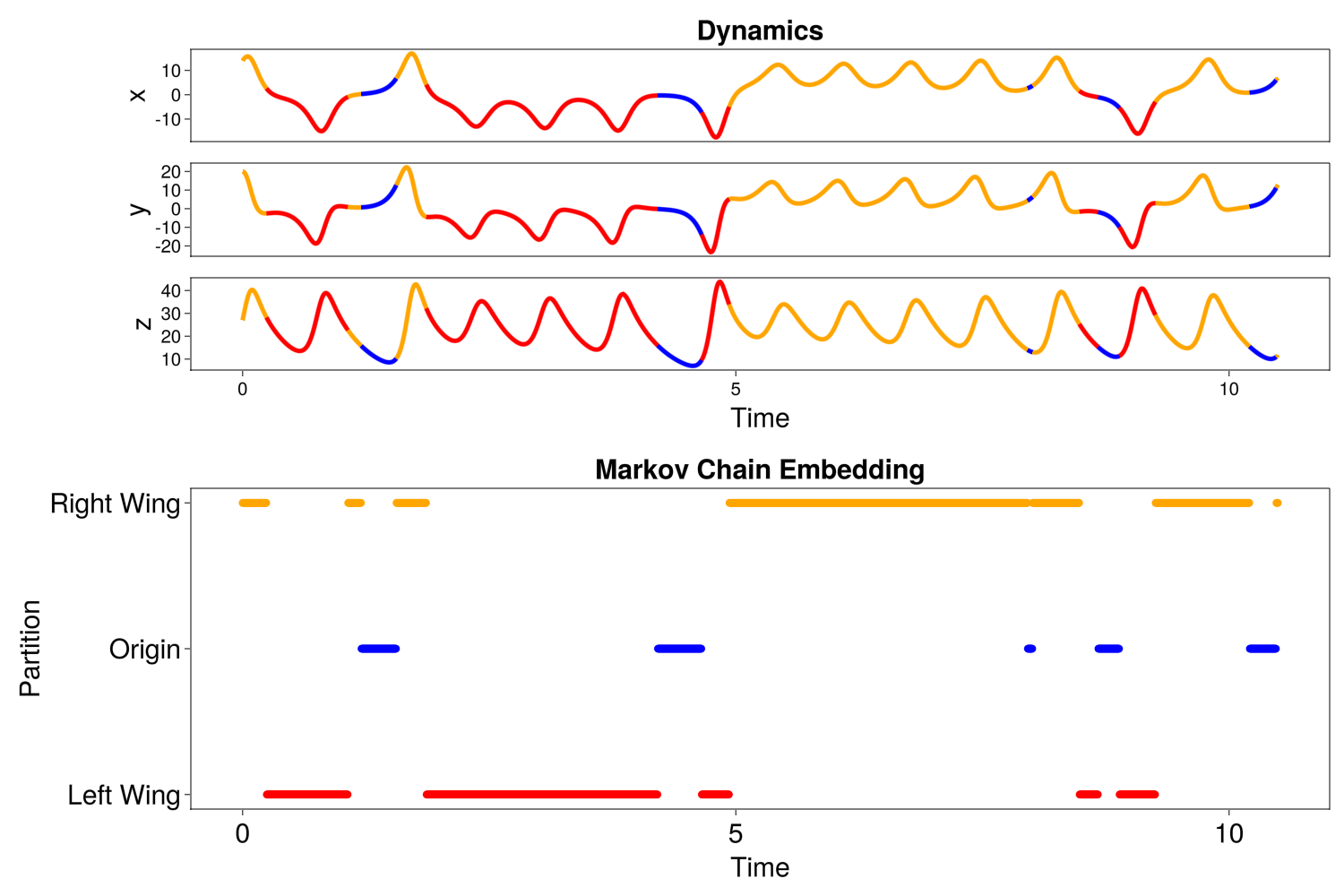}
\caption{Lorenz Fixed Point Partition Markov Chain Embedding. The dynamics of the $x, y, z$ variables are shown in the top 3 panels, and the associated embedding is shown in the bottom panel. As a dynamical trajectory moves through state space, it is labeled according to its proximity to the closest fixed point.}
\label{fig:lorenz_dynamics_embedding}
\end{center}
\end{figure}
\fi

We construct a time series from the Lorenz equations using a fourth-order Runge-Kutta time stepping scheme with time step  $\Delta t = 5 \times 10^{-3}$. We take the initial condition to be $(x(0), y(0), z(0)) = (14, 20, 27)$ and integrate to time $T = 10^5$, leading to $2 \times 10^7$ time snapshots. At each moment in time, we apply the embedding function to create a sequence of integers representing the partition dynamics\footnote{One can think of this as defining a symbol sequence.}. Figure \ref{fig:lorenz_dynamics_embedding} visualizes this process. 

From the sequence of integers, we apply the method from Section \ref{bayesian_construction} to construct the data-driven approximation to the generator with quantified uncertainty. For our prior distribution, we use an uninformative prior so that the mean of the random matrix agrees with the empirical construction from \ref{empirical_construction}. The mean for each entry of the generator (reported to two decimal places) is 
\begin{align}
\langle Q \rangle = 
\begin{bmatrix}
 -1.17   &  1.93   & 0.65  \\
  0.52   & -3.86  &  0.52 \\
  0.65   &  1.93  & -1.17
\end{bmatrix}
\end{align}
The apparent symmetry in the matrix results from the truncation to two decimal places and the abundance of data. In Appendix \ref{symmetries}, we show how to incorporate symmetries of the Lorenz equation and report ensemble mean statistics.

The utility of using a random matrix to represent uncertainty is summarized in Figure \ref{fig:lorenz_random_entries}. The distribution of each matrix entry for various subsets of time is displayed. Using fewer data (represented by a shorter gathering time, $T$) results in significant uncertainty for the entries. Additionally, using unconnected subsets of time demonstrates an apparent convergence of matrix entries.

\iffigure 
\begin{figure}
\begin{center}
\includegraphics[width=1.0\textwidth]{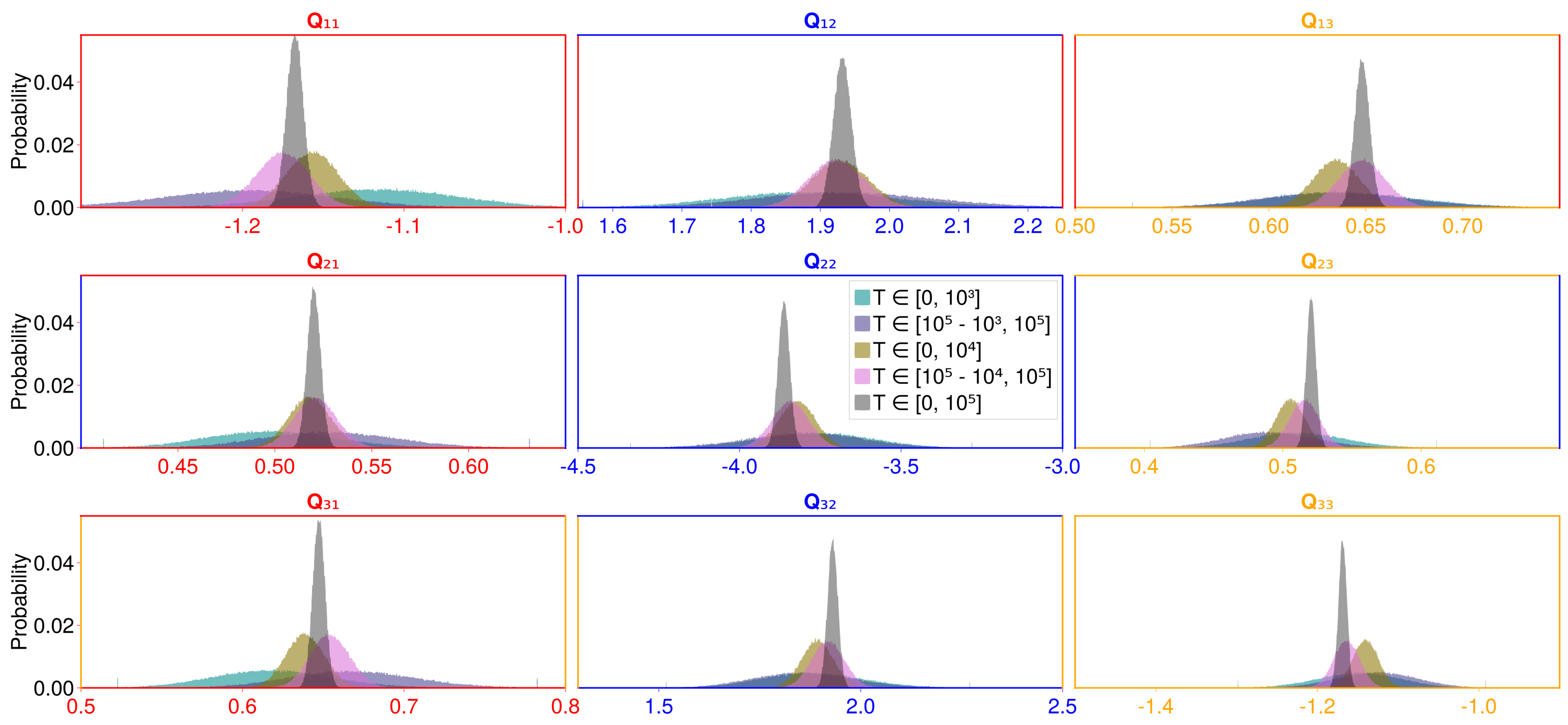}
\caption{Lorenz Fixed Point Partition Distributions of the Generator. The uncertainty estimates for the entries of the $3 \times 3$ generator are shown in the above figure. A one-to-one correspondence exists between the distributions in the panel and the matrix entries. The different colored distributions within a panel represent different estimates of the entries based on the amount of available data, here presented in terms of the simulation time of the Lorenz system. We see that as we increase the time interval of the simulation and thus have more data, we become more confident about the matrix entries. Furthermore, the distributional spreads overlap with one another.}
\label{fig:lorenz_random_entries}
\end{center}
\end{figure}
\fi

We are now in a position to calculate statistical quantities. For simplicity, we only report first, second, and third-order moments calculated from the mean value of the generator, $\langle Q \rangle$. The steady-state distribution of $\langle Q \rangle$, corresponding to eigenvalue $\lambda = 0$, is reported to two decimal places as
\begin{align}
[\mathbb{P}(\mathcal{M}_1) , \mathbb{P}(\mathcal{M}_2), \mathbb{P}(\mathcal{M}_3) ] \approx [0.44, 0.12, 0.44] 
\end{align}
from whence we calculate the steady state statistics for any observable using the approximations in Section \ref{ensemble_vs_time} and the Markov states $\pmb{\sigma}^n$ for $n = 1, 2,3$. Explicitly,
the ensemble average of the observables,
\begin{align}
g^1(\pmb{\mathscr{s}})=\mathscr{s}_{3} =  \mathscr{z}, \text{ } g^2(\pmb{\mathscr{s}})= (\mathscr{s}_{3})^2 = \mathscr{z}^2, \text{ or } g^3(\pmb{\mathscr{s}})=(\mathscr{s}_{1})^2\mathscr{s}_{3} =  \mathscr{x}^2 \mathscr{z} 
\end{align}
is approximated via Equation \ref{ensemble_average_approximation}, repeated here for convenience, 
\begin{align}
\langle g^j \rangle_E = g^j(\pmb{\sigma}^1) \mathbb{P}(\mathcal{M}_1)  + g^j(\pmb{\sigma}^2) \mathbb{P}(\mathcal{M}_2) + g^j(\pmb{\sigma}^3) \mathbb{P}(\mathcal{M}_3) \text{ for each } j
\end{align}
to yield
\begin{align}
\langle \mathscr{z} \rangle_E &\approx  27 \times 0.44 + 0  \times 0.12 + 27 \times 0.44 \approx 24 \\
\langle \mathscr{z}^2 \rangle_E &\approx 27^2 \times 0.44 + 0^2  \times 0.12 + 27^2 \times 0.44 \approx 642 \\
\langle \mathscr{x}^2 \mathscr{z} \rangle_E &\approx \left(-\sqrt{72} \right)^2 \times 27 \times 0.44 + 0^3 \times 0.12 + \left(\sqrt{72}\right)^2  \times 27  \times 0.44 \approx 1711 .
\end{align}

Table \ref{moment_table} shows the result from both the temporal and ensemble average. There is a correspondence for all averages, with the most significant discrepancy being those involving $y^2$ terms, for which the relative error is within $25\%$. The fixed points of a dynamical system are unique in that they satisfy all the same dynamical balances of a statistically steady state. Although we focused on moments, one can compare the statistics of any observable, e.g., 
\begin{align}
\langle \mathscr{z} \log(\mathscr{z}) \rangle_E \approx 78.4 \text{ and } \langle z \log(z) \rangle_T \approx 76.0,
\end{align}
where we used $z \log(z) \rightarrow 0$ as $z \rightarrow 0$. By symmetry one expects,
\begin{align}
\langle x \rangle = \langle y \rangle = \langle x z \rangle  = \langle y z \rangle = \langle y y y \rangle =  \langle xx y\rangle = \langle x yy \rangle = \langle x zz \rangle = \langle y zz \rangle = 0
\end{align}
but finite sampling effects prevent this from happening. As done in Appendix \ref{symmetries}, incorporating the symmetries allows ensemble calculations to achieve this to machine precision. 

\begin{table}
\centering
\begin{center}
\begin{tabular}{ c c c c c c c c c c}
  &  $\langle x \rangle$ & $\langle y \rangle$ & $\langle z \rangle$ & $\langle xx \rangle$ & $\langle xy \rangle$ & $\langle xz \rangle$ & $\langle yy \rangle$ & $\langle yz \rangle$ & $\langle zz \rangle$ \\ 
 \hline
 ensemble & -0.0 & -0.0 & 23.8 & 63.5 & 63.5 & -0.1 & 63.5 & -0.1 & 642.4 \\
 temporal & -0.0 & -0.0 & 23.5 & 62.8 & 62.8 & -0.2 & 81.2 & -0.2 & 628.9 \\
 \hline
   &   $\langle xxy \rangle$ & $\langle xxz \rangle$ & $\langle xyy \rangle$ & $\langle xyz \rangle$ & $\langle xzz \rangle$ & $\langle yyy \rangle$ & $\langle yyz \rangle$ & $\langle yzz \rangle$ & $\langle zzz \rangle$ \\
 \hline
 ensemble & -0.3 & 1713.2 & -0.3 & 1713.2 & -3.4 & -0.3 & 1713.2 & -3.4 & 17346.1 \\
 temporal & -0.4 & 1879.7 & -0.4 & 1677.2 & -4.1 & -0.4 & 1997.2 & -4.2 & 18446.3 \\
\end{tabular}
\end{center}
\caption{Empirical Moments of the Lorenz Attractor. A comparison between ensemble averaging and time averaging.} 
\label{moment_table}
\end{table}

In addition to containing information about steady-state distributions, the generator $Q$ provides temporal information: autocorrelations and the average holding time within a given partition.  
We show the autocorrelation of six observables,
\begin{align}
g^1(\pmb{\mathscr{s}}) &= \mathscr{x} \text{, } 
g^2(\pmb{\mathscr{s}}) = \mathscr{y} \text{, }  
g^3(\pmb{\mathscr{s}}) = \mathscr{z} \text{, }
g^4(\pmb{\mathscr{s}}) = 
\begin{cases}
    1 & \text{ if } \mathcal{E}(\pmb{\mathscr{s}}) = 1 
    \\
    0 & \text{ otherwise }
\end{cases}
\text{, } 
\\
g^5(\pmb{\mathscr{s}}) &= 
\begin{cases}
    1 & \text{ if } \mathscr{x} > 0
    \\
    -1 & \text{ if } \mathscr{x} < 0
    \\
    0 & \text{ otherwise}
\end{cases}
\text{, and }  g^6(\pmb{\mathscr{s}}) = 
\begin{cases}
    1 & \text{ if } \mathcal{E}(\pmb{\mathscr{s}}) = 2 
    \\
    0 & \text{ otherwise }
\end{cases}
\end{align}
in Figure \ref{fig:lorenz_autocorrelation}, which are calculated via Equations \ref{time_autocorrelation_approximation} and  \ref{ensemble_autocorrelation_approximation}, with appropriate modifications accounting for means and normalizing the height to one. Here we see both the success and limitations of the method at capturing autocorrelations. In general, the decorrelation of an observable is captured by the Markov model if it is approximately constant within a given partition, e.g., the observables $g^4$ and $g^5$. However, sometimes it is possible to do ``well", such as $g^1$ or $g^2$, despite not being constant within a region. We mention off-hand that $g^5$ is the numerical approximation to a Koopman mode (a left eigenvector of the generator $Q$) as induced by the partition. See Appendix \ref{Numerical_Koopman_Modes} for more details.
\iffigure
\begin{figure}
\begin{center}
\includegraphics[width=1.0\textwidth]{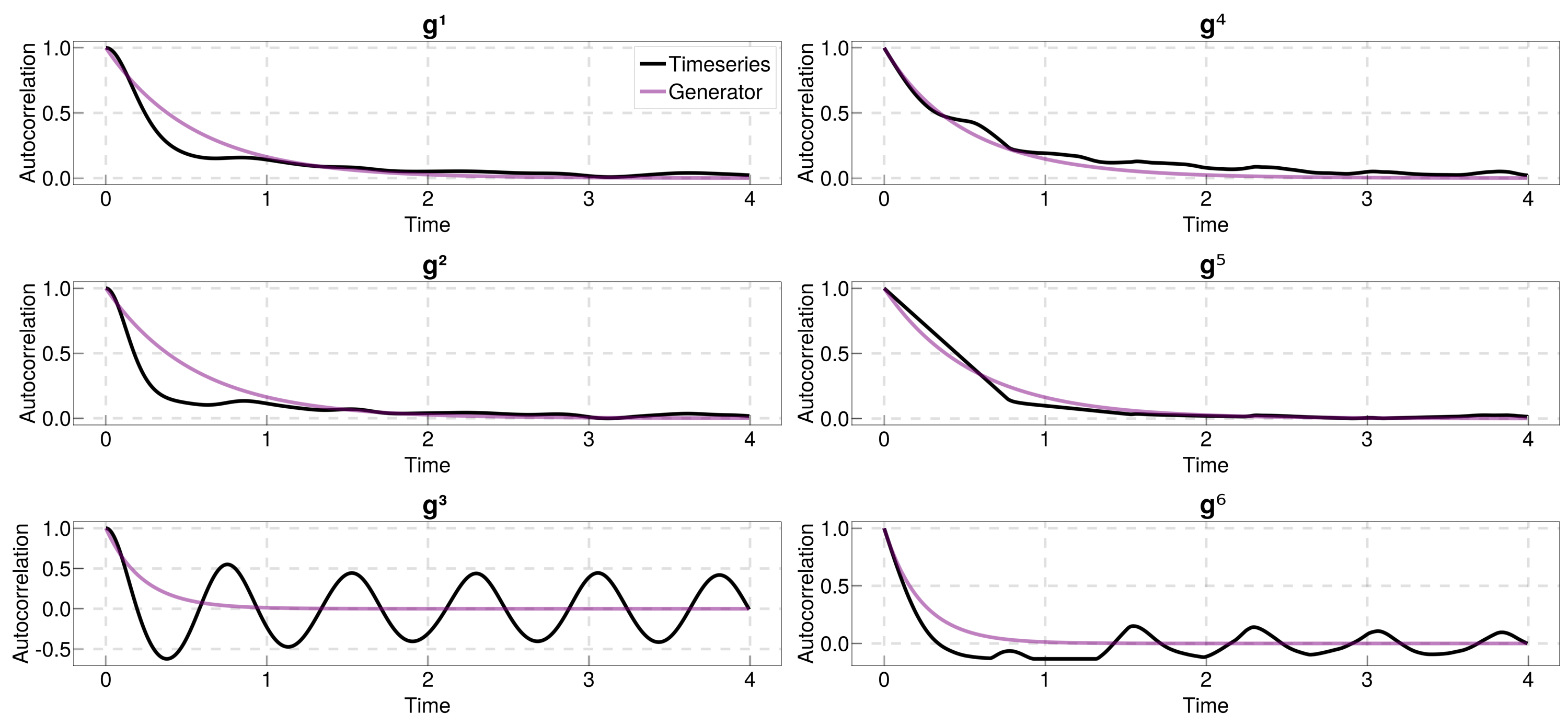}
\caption{Lorenz Autocorrelations Generator vs Timeseries. Six autocorrelations of observables are shown. The transparent purple line is calculated from the generator, and the black line is calculated from the time series. Even a coarse partition can capture observables $g^4$ and $g^5$ but struggles with oscillatory correlations. }
\label{fig:lorenz_autocorrelation}
\end{center}
\end{figure}
\fi

The inability to capture the autocorrelation of $g^6$, which is constant within $\mathcal{M}_2$, is partially due to the holding time distribution being far from exponentially distributed. To see this mode of failure, we plot the holding time distribution of the partitions in Figure \ref{fig:lorenz_holding_times}. We show several binning strategies of the distribution to demonstrate the ability of an exponential distribution to capture quantiles of the empirical holding time distribution.

Depending on the timescale of interest, the $\mathcal{M}_1$ and $\mathcal{M}_3$ partitions are approximately exponentially distributed, although they become fractal-like in terms of the distribution of holding times. In contrast, the holding time distribution of partition $\mathcal{M}_2$ is far from exponentially distributed upon refining the bins. Additionally, there is an inherent assumption in the construction of the generator that transition probabilities are independent of the amount of time spent in a particular subset of state space. A better statistical model would incorporate exit probabilities conditioned on the time spent in a partition. 

\iffigure
\begin{figure}
\begin{center}
\includegraphics[width=1.0\textwidth]{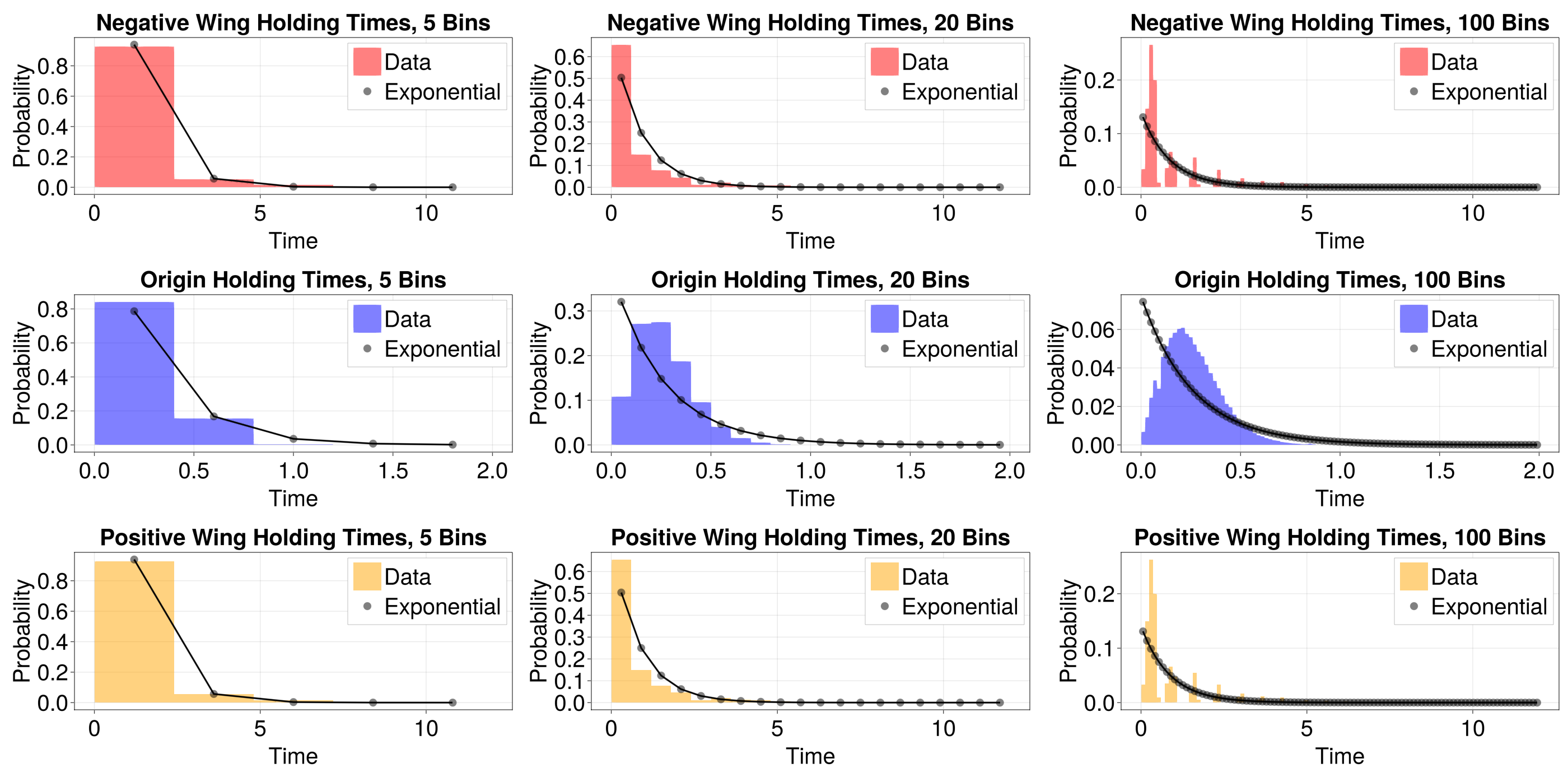}
\caption{Lorenz Fixed Point Partition Holding Times. An underlying assumption of using a generator for a given partitioning strategy is that the time spent in a state is exponentially distributed. Here we examine quantiles of the holding time distribution for a partition as given by the different binning numbers. The black dots correspond to the equivalent exponential distribution quantile, where the generator gives the rate parameter. }
\label{fig:lorenz_holding_times}
\end{center}
\end{figure}
\fi

Figure \ref{fig:lorenz_graph} summarizes the resulting statistical dynamics, where the generator and transition probabilities define a graph structure. The graph structure contains information about the topological connectivity between different regions of state space and the ``strength" of connectivity over different timescales of the dynamics as encapsulated by the transition probabilities. The generator itself is a generalized ``graph Laplacian" of the discrete system.

In the next section we move to a partial differential equation example where the same methodology applies, but is much more subtle in terms of what information can be extracted by using the method.

\iffigure
\begin{figure}
\begin{center}
\includegraphics[width=1.0\textwidth]{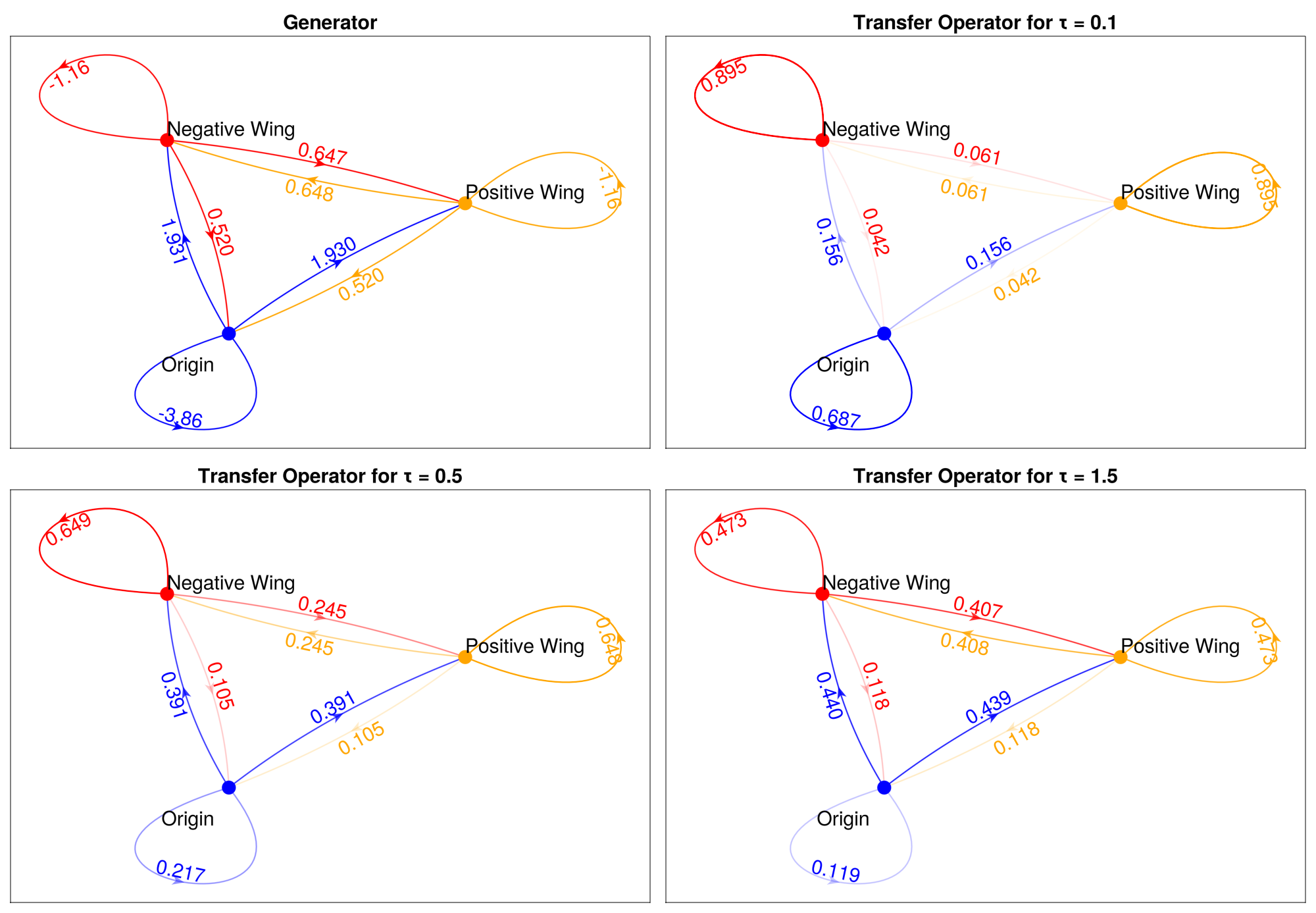}
\caption{Lorenz Fixed Point Graph.  The generator (top left) and transition probabilities over several timescales are visualized as a graph. The transition probabilities change depending on the timescale.  }
\label{fig:lorenz_graph}
\end{center}
\end{figure}
\fi

\subsection{Held-Suarez: Subtleties of High-Dimensional Discretizations}

We have seen how the methodology applies to a coarse discretization of the Lorenz statistics. We now apply the same methodology to the Held-Suarez atmospheric test case. In Figure \ref{fig:held_suarez_prognostic_fields}, we show a typical snapshot of the prognostic variables in the Held-Saurez simulation. The longitudinal velocity is the wind speed that flows in the east-west direction, and the meridional velocity flows in the north-south direction.

Due to the high dimensionality of the system, there are some subtleties to consider. For example, in the previous section, we saw that if a particular observable is uncorrelated with a given partition, then it is unlikely that the autocorrelations are well-captured by the generator. Furthermore, based on Equations \ref{ensemble_average_approximation} and \ref{ensemble_autocorrelation_approximation}, we expect only observables that are roughly constant within a partition will have faithful representations of their ensemble mean statistics.

However, both of these intuitions are not necessarily true. In the high-dimensional setting, we must distinguish between two classes of observables: Those highly correlated with a given partitioning strategy and those not. We rely on Monte-Carlo sampling to compute ensemble statistics for observables that are uncorrelated with a partitioning strategy. Effectively an observable is a random vector with respect to the partitions. 

Suppose the approximate generator yields a uniform distribution for the steady-state distribution. In that case, it is expected that, as long as the Markov states are ``independent" of one another, then one can do at least as well as Monte-Carlo sampling. If a partitioning strategy is well-correlated with an observable of interest, then we expect to do better than ``random" sampling of the Markov states.

\iffigure
\begin{figure}
\begin{center}
\includegraphics[width=1.0\textwidth]{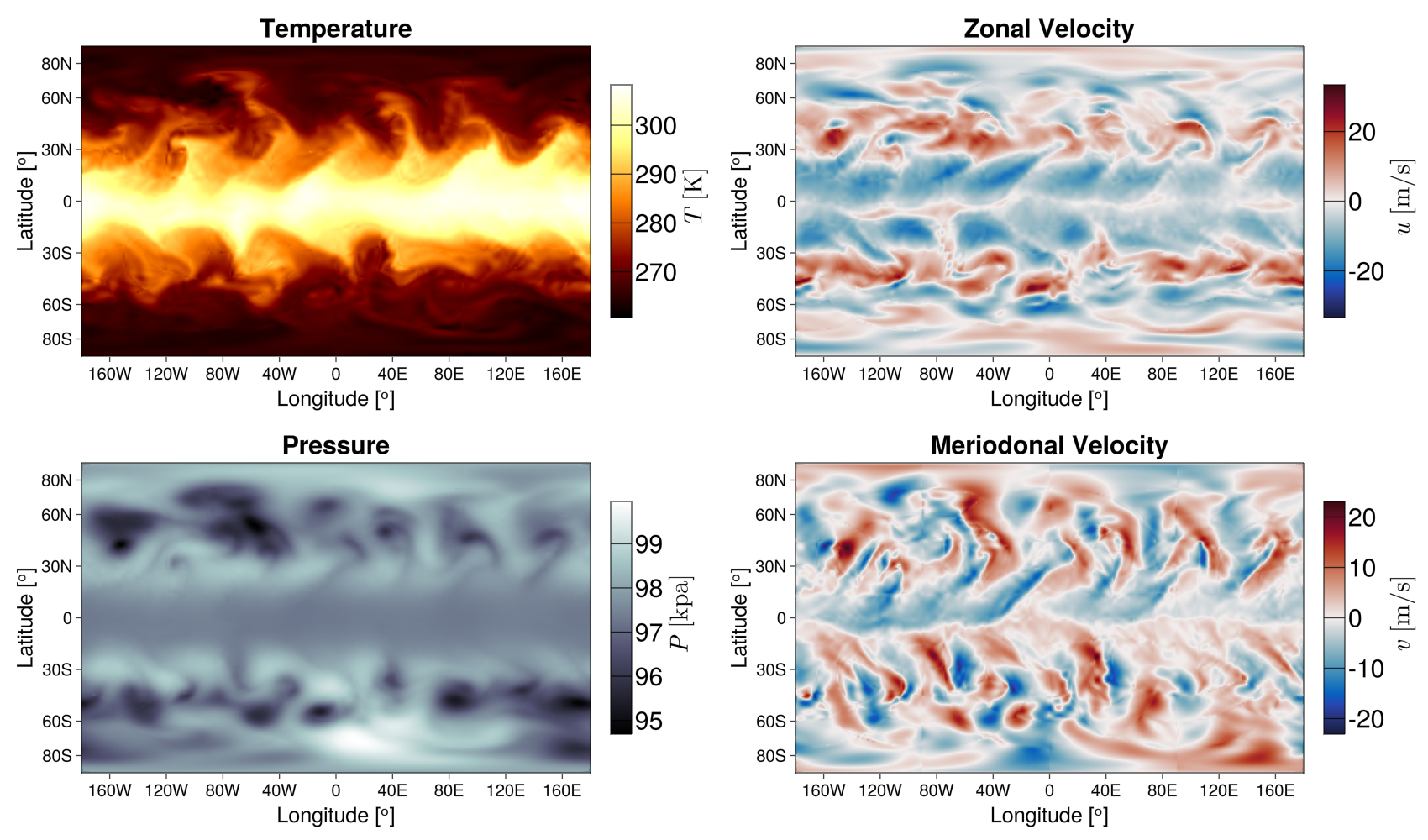}
\caption{Surface Fields of the Held-Suarez Atmospheric Test Case. We show surface temperature (top left), longitudinal velocity (top right), meridional velocity (bottom right), and pressure (bottom left).  }
\label{fig:held_suarez_prognostic_fields}
\end{center}
\end{figure}
\fi

For autocorrelations, a similar phenomenon occurs. An observable that is uncorrelated with the partitions is a random vector with respect to the partitions. In so far as there are many observables with similar autocorrelations, this strategy will do well to capture those observables. 

Of course, we do not wish to rely on luck or be at the mercy of the algorithm to target observables of interest. Thus in the high-dimensional setting, the chosen partitioning strategy, as encapsulated by the $\mathcal{E}$ function, is critical to targeting observables of interest. We go through two examples of choosing partitions. The first is similar to the previous section: define an appropriate distance function based on Markov states. The second is meant to target an observable of interest, temperature ``extremes" on the inner radius of the spherical shell (meant to represent heat waves at a fixed location). The latter is more akin to what is done in statistical mechanics, where one defines a ``macro-state", but in addition, we will pick out a few ``micro-states" corresponding to the macro state. In general, one can create partitions of the entire state space by partitioning according to one (or several) observables, as is commonly done when performing dimensionality reduction; however, we contend that we always want a representative state associated with a partition to calculate ensemble mean statistics and correlations associated with the total state space. 

The first partitioning strategy is chosen to provide insight into the topological structure of the turbulent attractor. The latter demonstrates that targeted partitioning strategies enable data-driven statistical modeling for observables of interest. 

\subsection{Held-Suarez: Distance Partitioning}

The first partitioning strategy starts with an initial simulation run to reach a turbulent state, as detailed in Appendix \ref{held_suarez}. A simulated ``day" is used as the unit of time, which corresponds to one rotation of the planet based on its angular velocity vector $\pmb{\Omega}$. In the atmosphere, the weather's decorrelation time is stated to be approximately two weeks. Thus, Markov states are gathered every 15 simulated days until 400 states have been accumulated. This choice corresponds to random samples of the turbulent attractor.

The embedding function $\mathcal{E}$, as before, corresponds to the index of the ``closest" Markov state. Our notion of ``close" is based on the distance function,
\begin{align}
d(\pmb{\mathscr{s}}^1, \pmb{\mathscr{s}}^2) &= \sqrt{ \int_\Omega d\pmb{x} \sum_{i} (\alpha_i)^{-2} \left( \pmb{\mathscr{s}}^1_{(\pmb{x}, i)} -  \pmb{\mathscr{s}}^2_{(\pmb{x}, i)} \right)^2 }
\end{align}
which is a weighted $L^2$ norm between the different fields of the system (so that we add fields together in a dimensionless way). The $\alpha_i$ are 
\begin{align}
\alpha_1 = 1.3 [kg / m^3], \alpha_2 = \alpha_3 = \alpha_4 = 60 [m/s], \alpha_5= 2.3 \times 10^6 [kg / (m s^2)].
\end{align}
In total the embedding function is 
\begin{align}
\mathcal{E}(\pmb{\mathscr{s}} ) &= n \text{ if }
d(\pmb{\mathscr{s}}, \pmb{\sigma}^n)   < d(\pmb{\mathscr{s}}, \pmb{\sigma}^m) \text{for each } m \neq n
\end{align}

We evolve the system an additional 200 simulated years and apply the embedding function $\mathcal{E}$ every $\Delta t = 0.03$ simulated days to the instantaneous state. This is a total of 2,000,000+ snapshots of time, but none of the snapshots are saved since this would've amounted to over 20 terabytes of data. The embedding function is applied ``on the fly" and only an integer sequence is recorded. The first 30 simulated days of this process is shown in Figure \ref{fig:held_suarez_embedding}. We have ordered the indices a posteriori so that the most probable partition is assigned index 1 and the least probable partition is assigned index 400. 

\iffigure
\begin{figure}
\begin{center}
\includegraphics[width=1.0\textwidth]{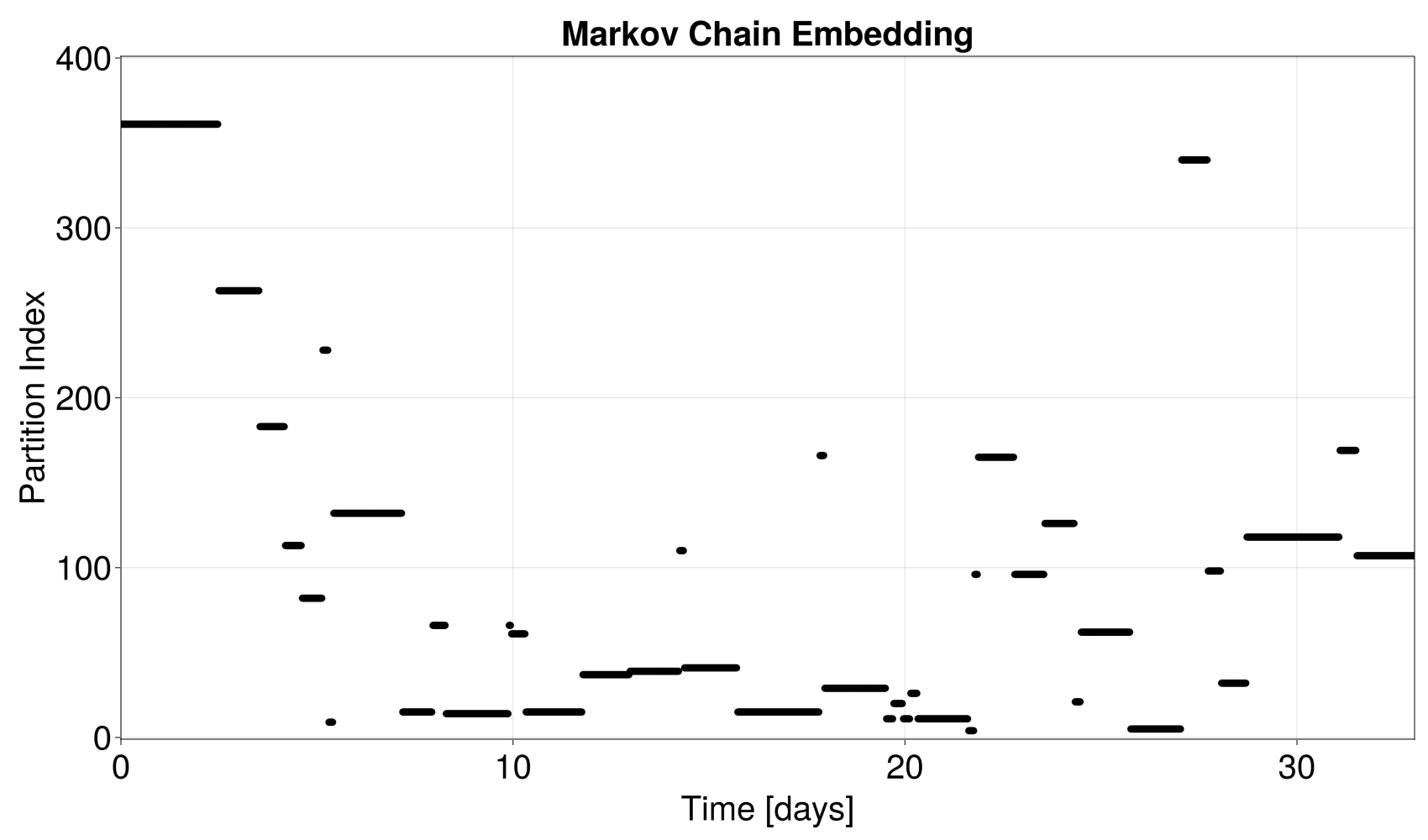}
\caption{Held Suarez Markov Chain Embedding. The dynamics are reduced to a sequence of integers. We order the indices by the steady-state probability of being with a partition so that index 1 corresponds to the most probable partition and index 400 to the least probable partition.}
\label{fig:held_suarez_embedding}
\end{center}
\end{figure}
\fi

For our prior distribution on the generator, we take $\alpha = 1$, and $\beta = \Delta t$, where $\Delta t$ is the sampling time interval for the time series. For each column of the matrix we use $\vec{\alpha} = 10^{-4} \pmb{1}$ where $\pmb{1}$ is the vector of all 1s. This prior distribution is interpreted as follows: If a partition is not observed, then it is assumed that the holding time is below the sampling threshold given by $\Delta t$ days. We take this precaution because it is not clear a-priori if every partition is revisited over a finite sampling period. That being said, 200 simulated years sufficed for revisiting every partition. The unobserved state is assumed to be sparsely connected and is reflected in the choice $\vec{\alpha} = 10^{-4} \pmb{1}$.

Displaying the mean and variance of a $400 \times 400$ random matrix is not particularly illuminating. Thus we summarize four properties of the mean generator in Figure \ref{fig:held_suarez_generator_properties}: the real part of the inverse 
eigenvalues, the steady state probability values associated with a partition, the connectivity of a given partition to every other partition, and the average holding time of a partition. The inverse eigenvalues' real part is associated with the slowest decaying autocorrelations of the system as captured by the partitioning strategy. We see that there is a clustering of eigenvalues around one simulated day. Furthermore, we see an apparent spectral gap between the first few eigenvalues and the bulk\footnote{It is not clear if there is a unique  limit upon refining a coarse-grained state space. This may imply the existence of both a continuous and discrete spectra in the limit of ever-refined partitions.}. The steady-state probability vector is not uniform (top left), and yet the amount of time spent in each state (bottom right) is roughly the same for each state. The reason for non-uniform probabilities is explicated by looking at the connectivity of a given partition (bottom left). The connectivity is defined as the empirical number of exits from or entrances to a given partition. We see that the more probable partitions are more connected to the rest of state space than the rest. The connectivity of a partition can be thought of as the effective dynamical predictability associated with a partition. For example, sufficiently sampled partitions of a periodic solution are only connected to one other partition since the future is precisely predictable from the past. 

Furthermore, the maximal connectivity is around 200, a number smaller than expected using 400 partitions that are based on distance in a 1,000,000+ dimensional space. It is precisely the reduced level of connectivity that frees the current partitioning strategy from the ``curse of dimensionality" present in Ulam's method, see \cite{Ulam1964}. The control ``volumes" are adapted to the shape of the turbulent attractor and dynamics guide the outflow of probability to only a subset of the myriad of faces associated with a box control volume. Ulam's method is a special case of the method in Section \ref{methodology} since one can take the centers of the boxes from Ulam's method as the Markov states and use an $L^{\infty}$ norm to compute distances between states. Thus, using the same boxes from Ulam's method would result in the same approximation to the generator. 

\iffigure
\begin{figure}
\begin{center}
\includegraphics[width=1.0\textwidth]{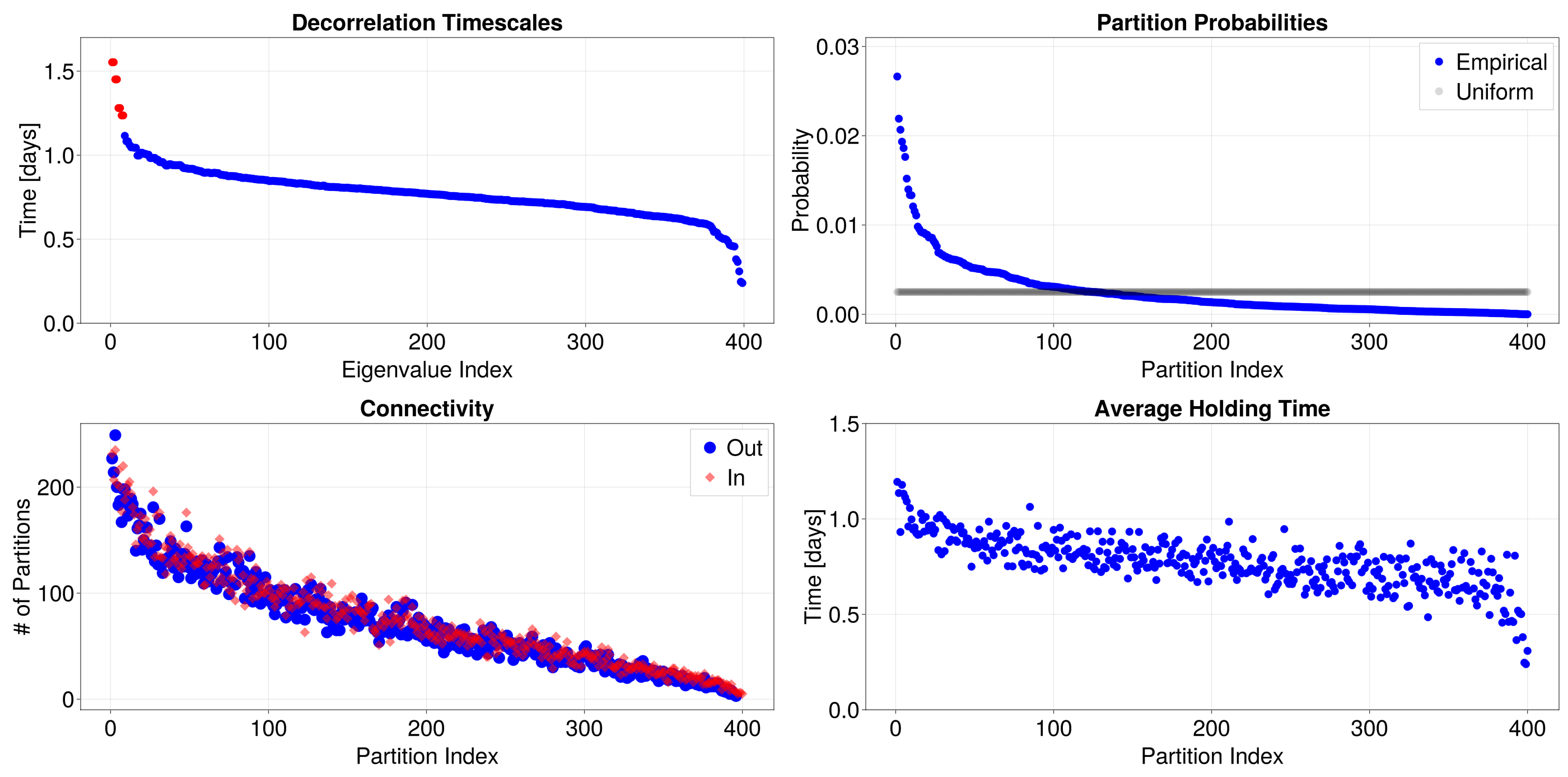}
\caption{Generator Properties. In the top left panel, we show the inverse real part of the eigenvalues of the generator, corresponding to the decorrelation time scales associated with the partitioning. The top right panel is the steady state probability associated with a partition. The bottom right is the holding time for a given partition. The bottom left summarizes the connectivity of a partition to other partitions based on the empirically observed transitions.}
\label{fig:held_suarez_generator_properties}
\end{center}
\end{figure}
\fi

A priori, there is no reason to expect any partition to be different from another partition, given that Markov states were sampled uniformly in time; however, Figure \ref{fig:held_suarez_generator_properties} suggests otherwise. The most probable regions of state space act as central hubs, connecting the various regions of state space together. These are perhaps associated with coherent structures such as fixed points or periodic orbits with few unstable directions. 

We have discussed the topological characteristics of the generator. Additional details on finite sampling effects and the holding time distributions are explored in Appendix \ref{further_hs_deetz}. In summary, holding times for the most probable states are approximately exponentially distributed, and there are significant uncertainties concerning the largest decorrelation timescale of the generator, but less-so for the other eigenvalues. We now move on to the calculation of statistical quantities.

As mentioned before, we distinguish between observables that are correlated with partitions and those that are not. The current partitioning strategy does not target an observable but does lead to emergent observables associated with slow and fast decorrelation timescales, i.e., global Koopman modes. Although not the focus of the present work, we discuss them in Appendix \ref{koopman_modes}. The focus for the text here is to calculate statistical quantities for observables that are \textbf{not} correlated with partitions. Later, in Section \ref{held_suarez_extreme_partition}, we construct a partition for a target observable.

We examine the histogram of the observable
\begin{align}
g(\pmb{\mathscr{s}}) &= \pmb{\mathscr{u}}_{\pmb{x}}\cdot \hat{\varphi}
\end{align}
where $\hat{\varphi}$ is the unit vector along the longitudinal direction and $\pmb{x}$ is a point on the inner shell (surface) at latitude $\theta = 35S$ and longitude $\varphi = 135E$, in Figure \ref{fig:held_suarez_steady_state}. We show two overlapping histograms. One histogram is calculated from the generator, and the other from gathering time series. The time series of the observable was accumulated over a 30 year timespan disjoint from the data used to construct the generator. The purple region is where the two histograms overlap, the red region is where the Markov model overpredicts the probability, and the blue region is where the Markov model underpredicts the probability. We show several bins, as before, to capture the notion of ``convergence in quantile". When we have as many bins as Markov states (400 bins in the present case), the delta function approximation begins to reveal itself. The height of the delta functions is associated with the steady-state probability distribution of the generator. The distribution was captured solely due to the choice of Markov states which constitute random samples of the turbulent attractor. The same level of fidelity is achieved by assuming a uniform distribution for the probability of the Markov states; there is no need to construct a generator to obtain this result. In the present context, this is a boon since these distributions are obtained ``for free". 

We further emphasize this point by calculating the mean for a continuum of observables. We use the longitudinal average of the longitudinal velocity field for each latitude, and each height, 
\begin{align}
g^{(\theta, r)}(\pmb{\mathscr{s}}) &=  \frac{1}{2\pi}\int_{0}^{2 \pi} (\pmb{\mathscr{u}}_{(\theta, \varphi, r) } \cdot \hat{\varphi}) d \varphi.
\end{align}
A fixed latitude $\theta$ and height $r$ constitute one observable and we expanded a position $\pmb{x} = \theta \hat{\theta} +  \varphi \hat{\varphi} +  r \hat{r}$ in terms of its components in a spherical basis. We calculate the ensemble and temporal mean for each observable in Figure \ref{fig:held_suarez_zonal_wind}. The temporal mean is gathered over three simulated years. In order to make a connection with how this field is usually visualized, we rescale the height of the axis according to the longitudinal average of pressure at the equator. This rescaling mimics the effect of using ``pressure coordinates" in the atmospheric literature. We see that the ensemble and temporal mean differ by less than two meters per second on the right half of the longitudinal wind ``butterfly" wing but is otherwise well-captured.   

\iffigure
\begin{figure}
\begin{center}
\includegraphics[width=1.0\textwidth]{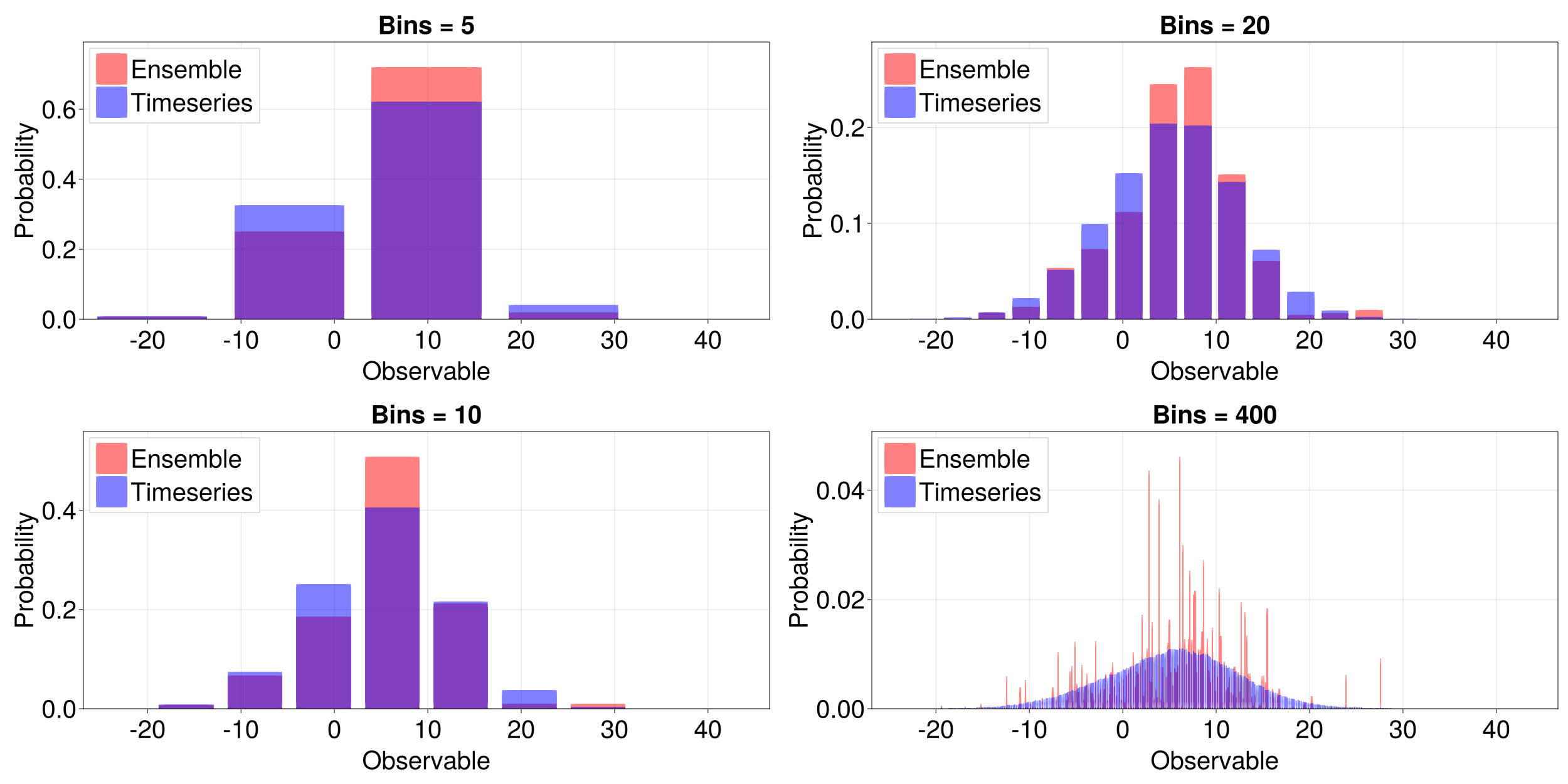}
\caption{Steady State Distribution of an Observable. Here we use the delta function approximation to the probability densities within a partition and look at the inferred distributions based on different coarse-grainings of the distribution. The overlap region is in purple; red bars correspond to ``overpredicting" probabilities, and blue bars correspond to ``underpredicting" probabilities. The temporal and ensemble means are $5.3$ and $5.7$ respectively. The temporal and ensemble standard deviations are $7.1$ and $6.5$ respectively. }
\label{fig:held_suarez_steady_state}
\end{center}
\end{figure}
\fi

\iffigure
\begin{figure}
\begin{center}
\includegraphics[width=1.0\textwidth]{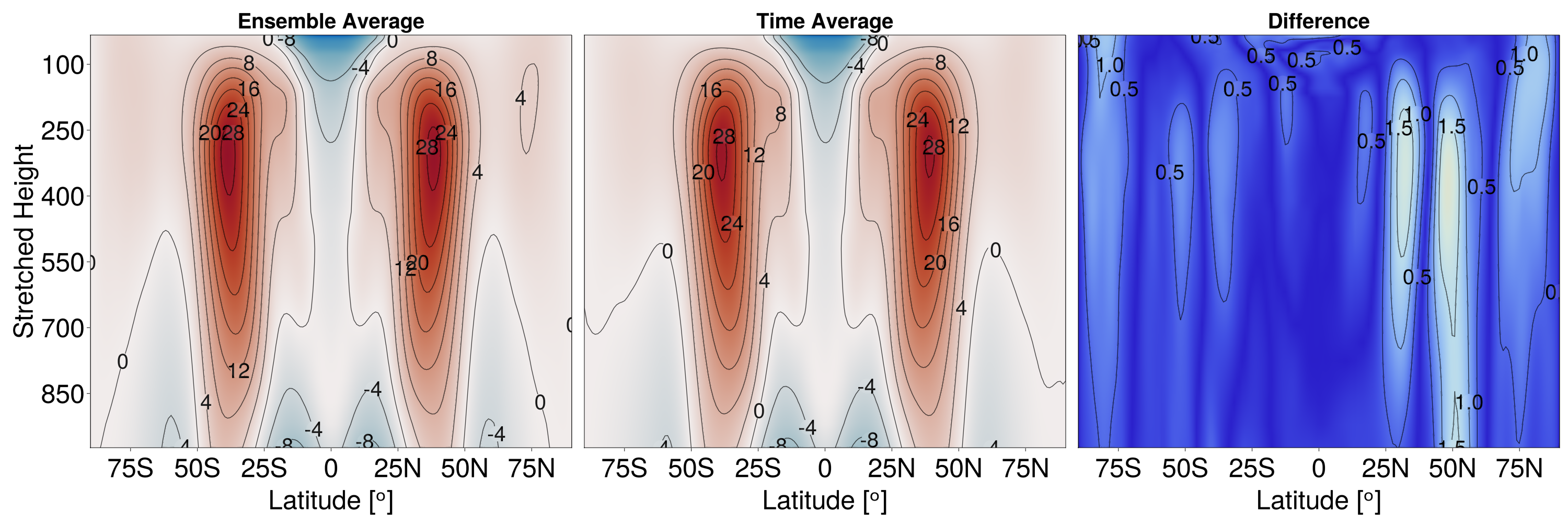}
\caption{Mean Value for a Continuum of Observables. The ensemble average (left) and temporal average (middle) longitudinal mean longitudinal wind display a mean for a  continuum of observables. The plot farthest to the right shows the point-wise absolute difference between the two means.}
\label{fig:held_suarez_zonal_wind}
\end{center}
\end{figure}
\fi

The autocorrelation of four observables, 
\begin{align}
g^1(\pmb{\mathscr{s}}) =  \varrho_{\pmb{x}}  \text{ , } 
g^2(\pmb{\mathscr{s}}) =  \pmb{\mathscr{u}}_{\pmb{x}} \cdot \hat{\varphi} \text{ , } 
g^3(\pmb{\mathscr{s}}) =  \pmb{\mathscr{u}}_{\pmb{x}} \cdot \hat{\theta} \text{ , and } 
g^4(\pmb{\mathscr{s}}) =  \mathscr{T}_{\pmb{x}} \text{ , } 
\end{align}
are shown in Figure \ref{fig:held_suarez_autocovariance} for the same position $\pmb{x}$ as before. The variable $\mathscr{T}_{\pmb{x}}$ is the temperature at the same point and defined through the relation, 
\begin{align}
\label{temperature_definition}
\mathscr{T}_{\pmb{x}} &\equiv \frac{\gamma - 1}{ R_d \varrho_{\pmb{x}} }(\varrho e_{\pmb{x}} - \frac{1}{2} \varrho_{\pmb{x}}\|\pmb{\mathscr{u}}_{\pmb{x}}\|^2 -  \varrho_{\pmb{x}} \Phi_{\pmb{x}}) 
\end{align}
where $R_d = 287$ is the ideal gas constant, $\gamma = 1.4$ is the specific heat ratio of air, and $\Phi$ is the geopotential.
We show the empirically obtained autocorrelation from the time series in blue and the generator in purple. Since most of the eigenvalues of the generator cluster around one day, a random vector is likely to produce a decorrelation time of one day. This partition is a poor approximation for observable $g^3$, but not-so for the other variables. In principle, there are infinitely many observables whose decorrelation times are well approximated by a random vector and infinitely many other poorly approximated observables. 

\iffigure
\begin{figure}
\begin{center}
\includegraphics[width=1.0\textwidth]{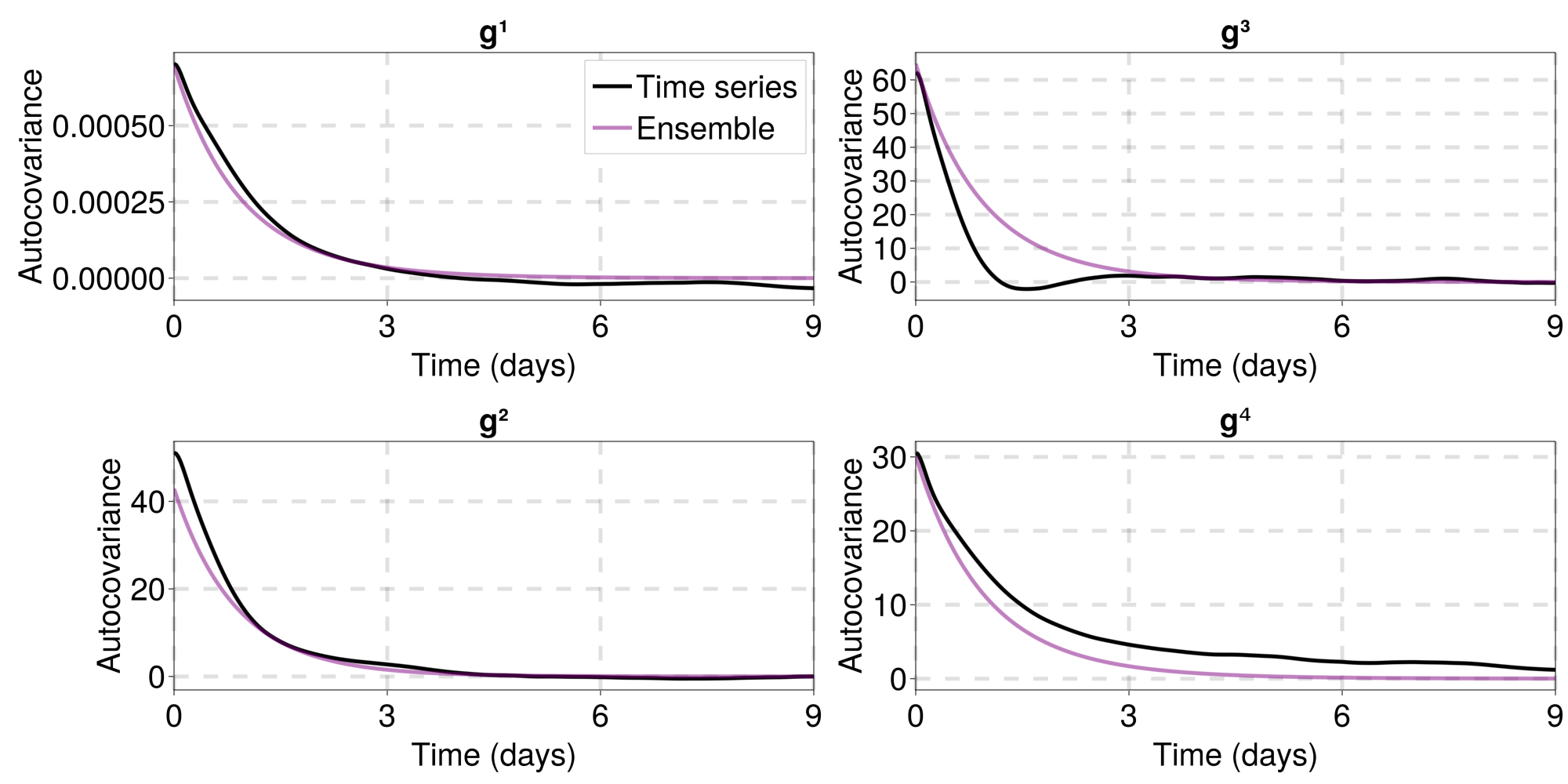}
\caption{Autocovariance for Several Observables in the Held-Suarez Atmospheric Test. The autocovariance for several observables based on the time series (black) and generator (purple) are shown. The observables are uncorrelated with the partition; thus, the autocorrelation predicted from the Markov model is similar for all four cases.}
\label{fig:held_suarez_autocovariance}
\end{center}
\end{figure}
\fi

Thus, the following section shows what happens when a particular observable is targeted. 

\subsection{Held-Suarez: Extreme Statistics Partitioning}
\label{held_suarez_extreme_partition}

We now partition the turbulent attractor in a different way to target statistics of a particular observable: Temperature extremes at particular point in the domain. In particular,
\begin{align}
g(\pmb{\mathscr{s}}) &= 
\begin{cases}
    1 & \text{ if } \mathscr{T}_{\pmb{x}} > 290 [K] 
    \\
    0 & \text{ otherwise}
\end{cases}
\end{align}
Here  $\pmb{x}$ is a point on the inner shell of the sphere at latitude-longitude $(\theta = \text{35S}, \varphi = \text{135E})$. We use the script  $\mathscr{T}$ for temperature in analogy to the previous notation. The choice of 290 [K] came from the 95\% quantile of temperature at that point over a short simulation run.

We gather the Markov states by first partitioning an arbitrary state into two classifications: $g(\pmb{\mathscr{s}}) = 1$ and $g(\pmb{\mathscr{s}}) = 0$. The former is representative of an ``extreme state" and the latter of a ``benign state." We then gather ten representative ``extreme" states and ninety representative benign states. Specifically, a simulation is run, and the states are checked every two weeks. We apply the observable (i.e. classifier) $g$ to determine whether or not the state is extreme. The process is continued until at least ten extreme states, and ninety benign states are gathered. We only keep one hundred total states. Thus, any extra states are discarded. The extreme states are assigned indices 1-10, while the benign states are assigned indices 11-100.

With these Markov states in place, the embedding function is defined as follows 
\begin{align}
\mathcal{E}(\pmb{\mathscr{s}}) 
&= 
\begin{cases}
n & \text{ if } g(\pmb{\mathscr{s}}) = 1 \text{ and }
d(\pmb{\mathscr{s}}, \pmb{\sigma}^n)   < d(\pmb{\mathscr{s}}, \pmb{\sigma}^m) \text{for each } m \neq n \text{ where } m,n \in \{1, ..., 10\}
\\
n & \text{ if } g(\pmb{\mathscr{s}}) = 0 \text{ and }
d(\pmb{\mathscr{s}}, \pmb{\sigma}^n)   < d(\pmb{\mathscr{s}}, \pmb{\sigma}^m) \text{for each } m \neq n \text{ where } m,n \in \{11, ..., 100\}
\end{cases}
\end{align}
That is to say, we first classify the state according to the observable $g$, and then we calculate the closest Markov state within the respective category. Finally, we run the model for 100 simulated years and construct the Markov chain embedding.

We first focus on the holding time distribution of being in a partition associated with an extreme event. This distribution is taken as a proxy for the duration of a heatwave. Given that we have ten possible states corresponding to an extreme event, we also account for transitions between states within an extreme event duration. Heuristically, transitions between different global states during an extreme event are rare since the duration is short compared to the holding time of being in a state.

Nevertheless, they do occur in this simulation. Figure \ref{fig:held_suarez_extreme_graph} summarizes the transition pathways between partitions associated with extreme states. In this figure, states 11-100 have been lumped together as a single state, and the graph structure of the transition pathways is shown. The transparency of the red lines corresponds to the probability of transitioning between the different extreme state partitions, and the blue lines correspond to the transition probability of leaving an extreme state. The graph structure shows that the global state changes when an extreme event occurs at a particular location on the globe.

Furthermore, it reveals that an extreme state has many ``microstates" corresponding to the macrostate (defined by the partition induced from $g(\pmb{\mathscr{s}}) = 1$), and there are non-zero transitions between the ``microstates" during a given macrostate configuration. This nuance partially explains the complexity of an extreme event prediction. 

\iffigure
\begin{figure}
\begin{center}
\includegraphics[width=0.75\textwidth]{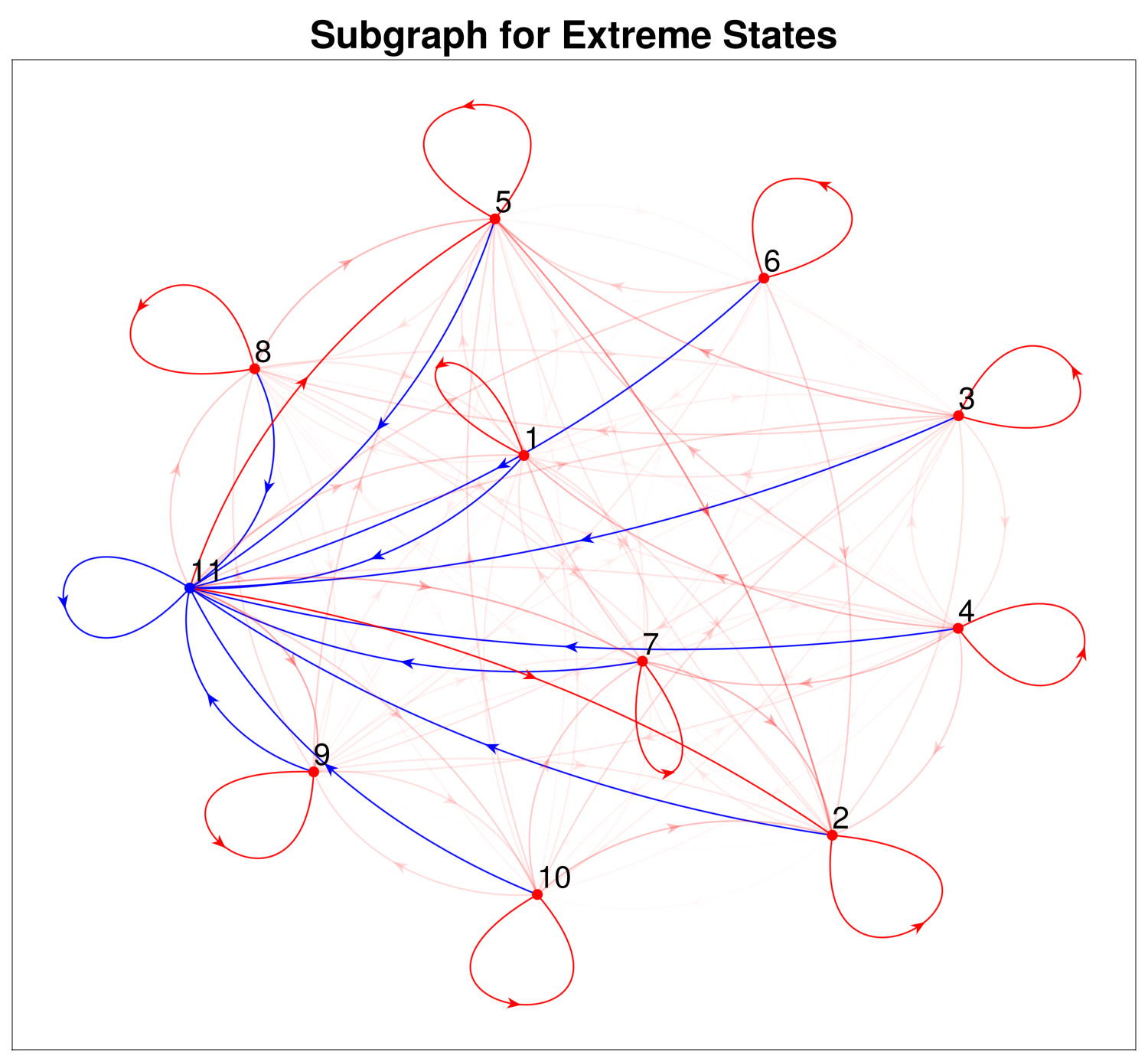}
\caption{Network Structure of Extreme Transitions.  Eleven states are shown where 1-10 correspond to an extreme state, and 11 correspond to the other 90 states, lumped together as a single node for visualization purposes. The blue lines correspond to transitions to a benign state and the red lines as transitions to extreme states. The opacity of the lines is proportional to the probability of transitioning between states. There exist transitions between extreme states.}
\label{fig:held_suarez_extreme_graph}
\end{center}
\end{figure}
\fi

The holding time distribution of being in an extreme state (as calculated by the Markov embedding) accounts for transitions between the different states. Furthermore, we gather statistics from the temperature observable at a disjoint set in time and show its holding time distribution in Figure \ref{fig:held_suarez_extreme_holding_time}. The Markov state representation shows that the holding time distribution is well-captured.

The presence of an extreme state can be viewed as an exit time problem from the point of view of stochastic processes. An extreme event corresponds to a particular subset of state space, characterized by here ten partitions. The average amount of time spent in an extreme state must incorporate the transitions within the duration of an extreme event. 

\iffigure
\begin{figure}
\begin{center}
\includegraphics[width=1.0\textwidth]{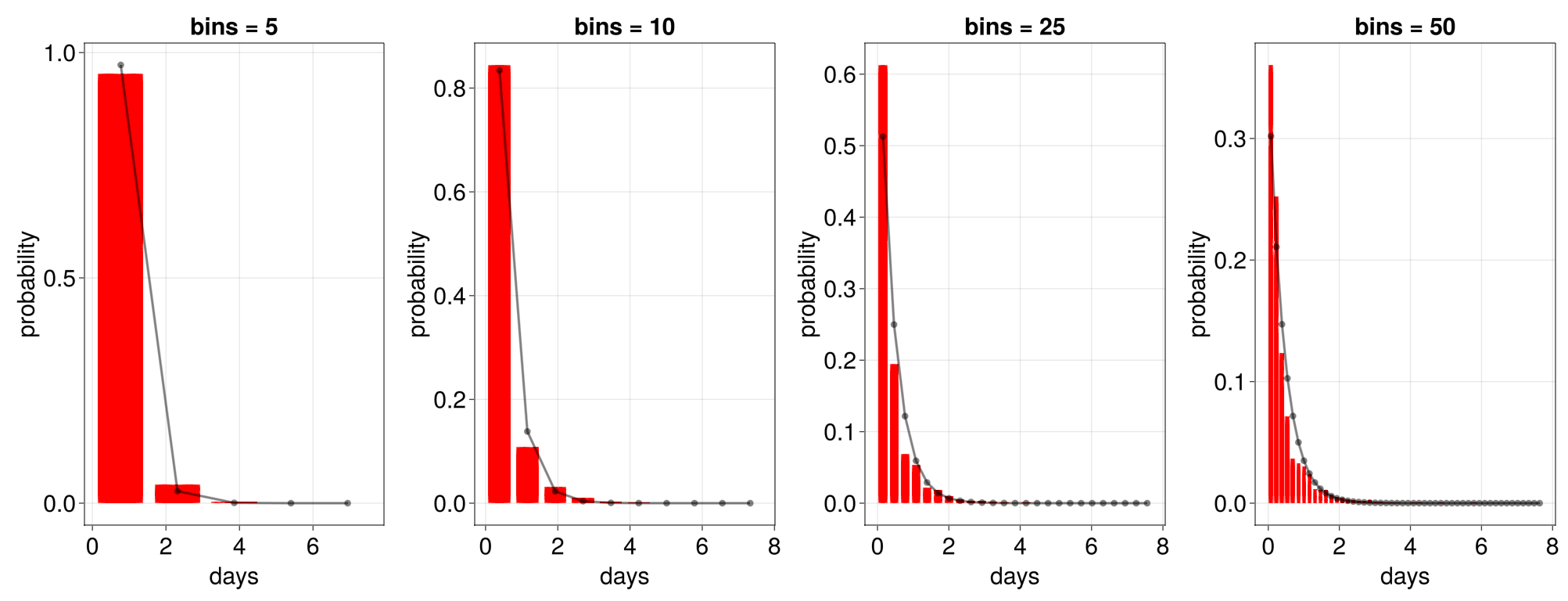}
\caption{Held Suarez Holding Time Extreme. Several quantiles for the duration of extreme states, as calculated from the time series, are shown in red. For simplicity, we show the exponential distribution holding time as black dots where the decorrelation time is approximately $1/2$ a day as calculated by looking at the holding time for a single extreme state partition.}
\label{fig:held_suarez_extreme_holding_time}
\end{center}
\end{figure}
\fi

We also compare the temperature distribution at $\pmb{x}$ as calculated by the 400-state system, the 100-state system, and the time series in Figure \ref{fig:held_suarez_observable_comparison}. In particular, we see that the 100-state system better captures the 95\% tail distribution. Thus selecting a partitioning strategy that targets an observable of interest is feasible and is of increased fidelity compared to the naive generic partitioning. This procedure is equivalent to local grid refinement from numerical methods.

\iffigure
\begin{figure}
\begin{center}
\includegraphics[width=1.0\textwidth]{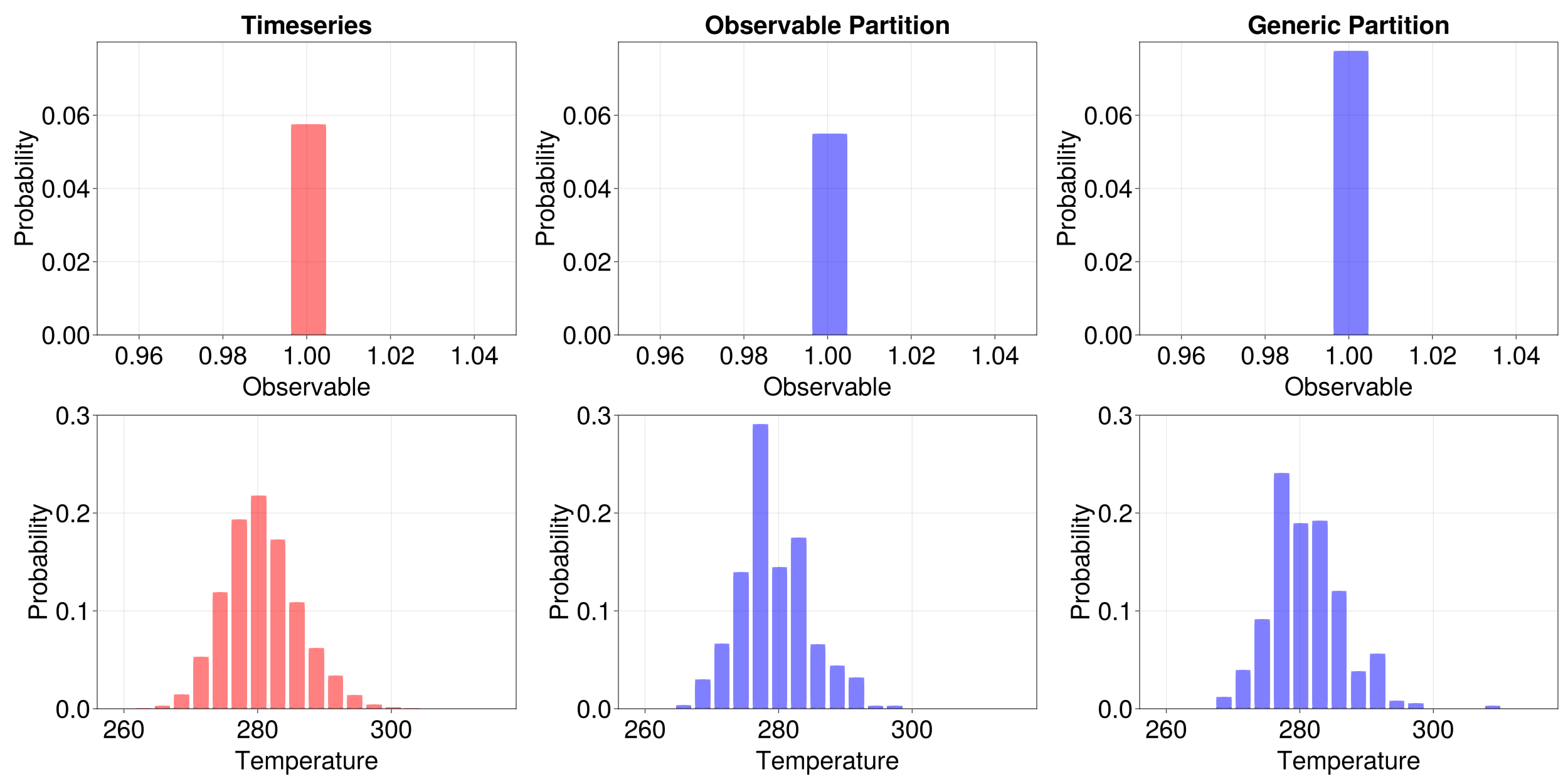}
\caption{Held Suarez Observable Comparison. We show the effect of grid refinement on an observable of interest. The refinement strategy better captures the tail probability quantile than the generic 400-state partition (top three panels). The temperature distribution is shown in the bottom three panels as a point of comparison.}
\label{fig:held_suarez_observable_comparison}
\end{center}
\end{figure}
\fi

As a final comment for this section, the choice of partition does not need to be binary. As one runs a simulation, the same Markov states can be used to compute several different partitioning strategies simultaneously. If partitioning strategies make use of independent observables, then it is natural to construct tensor product generators. We did not pursue any of these strategies here.

\section{Conclusion}
\label{discussion}
In summary, we have done three things 
\begin{enumerate}
    \item Section \ref{theory}: Reviewed and extended a theoretical formulation for transforming a dynamical system into a continuous time Markov process with finite state space.
    \item Section \ref{methodology}: Developed a Bayesian stream-based data-driven algorithm for constructing the generator of a continuous-time Markov process with finite state space.
    \item Section \ref{applications}: Applied the methodology to two systems: The continuity equations for the Lorenz system and the compressible Euler equations. 
\end{enumerate}

We have seen that many statistics can be captured even with a coarse discretization. In the Lorenz case, we used the fixed points of the dynamical system as both the Markov states and to anchor the partitioning strategy. The three states sufficed to capture mean and second moments. Furthermore, even some autocorrelations and residency times were well-captured with the coarse discretization, depending on the timescale of interest. 

Calculations are subtle in the high-dimensional setting. Observables uncorrelated with the partitioning strategy still give sensible answers for steady-state statistics due to Monte Carlo sampling. Furthermore, the autocovariance could also be captured for uncorrelated random variables, albeit not all possible observables. When the partitioning strategy is chosen to coincide with an observable of interest, the residency times and statistics were well captured. 

Taken together, we see that the most critical component in the statistical representation of a system is the choice of partitioning strategy. Future directions necessitate the development of novel partitioning strategies. For example, partitioning according to modal amplitudes given by Dynamic Mode Decomposition or using machine-learning methods such as auto-encoders to reduce the dimensionality of state space. It is likely that consistency between the Koopman and Perron-Frobenius operators would yield the greatest improvement. Incorporating partial temporal coherence in the Markov state partitioning also seems promising\footnote{For example, in the Held-Suarez case choosing Markov states that are one day apart for a month, then skipping a few months, and repeatedly gathering Markov states.}.

As a practical matter the partitioning strategy should be computational feasible. Using a tree structure to partition state space would greatly ameliorate the computational burden at the cost of more up-front memory. For example, using a tree structure and a binary classification would yield $2^N$ partitions, where $2^{N+1} - 2$ fields would have to be stored, and $2N$ evaluations of the classification function would have to be applied. Compression of the Markov states also becomes important in which case both symmetries and lower precision arithmetic should be used.

Since the method has been formulated as a numerical discretization, there are straightforward generalizations to consider. For example, in addition to discretizing space using a finite volume method one can discretize time using a Discontinuous Galerkin method. In this way time trajectories are represented as piecewise polynomial as opposed to piecewise constant. Furthermore the flux of probability to a different region of state space would now (in discrete time) depend on the history. 

A more radical departure from the methods proposed here is to use generative models, similar to \cite{ho2020denoising}, to represent distributions within a partition. Partitions of state space may be more amenable to representation than the entirety of the manifold. Furthermore, the use of nonlinear models for the generator to account for all the different physical features that one must assign ``attention", could yield a better overall representation, for example \cite{Vaswani2017}.

The primary reason for undertaking the perspective in this manuscript was to gain a foothold in understanding climate change from an operator-theoretic approach, similar to \cite{Froyland2018}. Climate change is often characterized as ``statistics changing over time" and thus requires a precise definition. We focused on a high-dimensional measure that is invariant with respect to time. This trait is not valid for the climate system, whose statistics are non-stationary. The predominant signal for a ``stationary" climate is not stationary but rather time-periodic due to the diurnal and seasonal cycles. Thus the first simplification is to consider a generator whose entries are periodic functions of time and whose Markov states are also periodic functions of time, see \cite{Wang2015} for similar considerations in molecular dynamics. Climate change is then characterized as deviations from this time-periodic (high-dimensional) flow.

\backsection[Supplementary data]{\label{SupMat}Supplementary material are available at \url{https://github.com/sandreza/MarkovChainHammer.jl}.}

\backsection[Acknowledgements]{Tobias Bischoff, Keaton Burns, Glenn Flierl, Raf Ferrari, Ludovico, Peter, MG, Fabri, Matthew, Simone, Pedram, Predrag. The author would like to thank the 2022 Geophysical Fluid Dynamics Program, where much of this work was completed. }

\backsection[Funding]{This work is supported by the generosity of Eric and Wendy Schmidt by recommendation of the Schmidt Futures program. The Geophysical Fluid Dynamics Program is supported by
the National Science Foundation, United States, and the Office of Naval
Research, United States.}

\backsection[Declaration of interests]{The author reports no conflict of interest.}

\appendix

\section{Global Koopman Modes}
\label{koopman_modes}

We do not ask, ``Can we predict an observable of interest?" but rather, ``What can we predict?". The latter question is an emergent property of the system and captured by the Koopman modes of the underlying system. Those Koopman modes whose decorrelation timescales are long-lived constitute the most predictable features of the system over long time scales.

Koopman modes are observables as well as left-eigenvectors of the transition probability operator $\mathcal{T}^\tau$. For example, if $g_\lambda$ is a left eigenvector of $\mathcal{T}^\tau$ with eigenvalue $e^{\lambda \tau}$ then we have the following 
\begin{align}
    R_E(g_\lambda, \tau) 
    &= \int_{\mathcal{M}} d\pmb{\mathscr{s}}' g_\lambda(\pmb{\mathscr{s}}' )   \mathscr{P}( \pmb{\mathscr{s}}' )\left[ \int_{\mathcal{M}} d\pmb{\mathscr{s}} g_\lambda(\pmb{\mathscr{s}}) \mathscr{T}^\tau \delta ( \pmb{\mathscr{s}} - \pmb{\mathscr{s}}') \right] \\ 
    &= \int_{\mathcal{M}} d\pmb{\mathscr{s}}' g_\lambda(\pmb{\mathscr{s}}' )   \mathscr{P}( \pmb{\mathscr{s}}' )\left[ \int_{\mathcal{M}} d\pmb{\mathscr{s}} g_\lambda(\pmb{\mathscr{s}}) e^{\lambda \tau} \delta ( \pmb{\mathscr{s}} - \pmb{\mathscr{s}}') \right] \\ 
    &= e^{\lambda \tau} \int_{\mathcal{M}} d\pmb{\mathscr{s}}' g_\lambda(\pmb{\mathscr{s}}' )^2   \mathscr{P}( \pmb{\mathscr{s}}' ) \\
    &= e^{\lambda \tau} \langle g_\lambda ^2 \rangle_E
\end{align}
Thus the most useful Koopman modes, from a predictability standpoint, are those such that decorrelate slowly in time, i.e. $\text{real}(\lambda) \approx 0$, but additionally have an oscillatory component so that the ratio $\text{real}(\lambda)/\text{imaginary}(\lambda) \approx 0$ holds.

If $\text{real}(\lambda) = 0$ on a chaotic attractor, then we expect this to be the ``trivial" observable\footnote{The presence of pure-imaginary eigenvalues would imply the existence of observables that are predictable for arbitrary times in the future on a chaotic attractor, which is incommensurate with the definition of an SRB measure.} $g_\lambda(\pmb{\mathscr{s}}) = c$ for a constant $c$. Otherwise, we expect that $\text{real}(\lambda) < 0$ for all eigenvalues corresponding to the transfer operator, i.e., we expect that all non-trivial observables will eventually decorrelate. This implies $\langle g_\lambda \rangle_E = 0$ since 
\begin{align}
\langle g_\lambda \rangle_E  &= \int_{\mathcal{M}} d \pmb{\mathscr{s}} g_\lambda( \pmb{\mathscr{s}} ) \mathscr{P}(\pmb{\mathscr{s}} ) = \lim_{\tau \rightarrow \infty} \int_{\mathcal{M}} d \pmb{\mathscr{s}} g_\lambda( \pmb{\mathscr{s}}) \mathscr{T}^\tau \delta ( \pmb{\mathscr{s}} - \pmb{\mathscr{s}}')  = \lim_{\tau \rightarrow \infty}  e^{\lambda \tau} g_\lambda(\pmb{\mathscr{s}}') = 0
\end{align}
where $\pmb{\mathscr{s}}'$ is an arbitrary state on the attractor $\mathcal{M}$. 

The following four statements about a Koopman mode $g_{\lambda}$ can't hold simultaneously
\begin{itemize}
    \item The Koopman mode satisfies the relation $g_\lambda(\pmb{s}(t+\tau)) = e^{\lambda \tau} g_\lambda(\pmb{s}(t))$
    \item The Koopman mode $g_\lambda$ is a continuous function of state space
    \item There exist an arbitrary number of near recurrences on the dynamical trajectory
    \item The eigenvalue associated with the Koopman mode satisfies $\text{real}(\lambda) < 0$.
\end{itemize}
The proof is as follows. Suppose that all four criteria are satisfied. Let $\pmb{\mathscr{s}}'$ be a near-recurrance of $\pmb{\mathscr{s}}$ some time $\tau$ in the future so that $\| \pmb{\mathscr{s}} - \pmb{\mathscr{s}}' \| < \epsilon$ for some norm. Continuity of $g_\lambda$ with respect to the norm implies 
\begin{align}
| g_\lambda( \pmb{\mathscr{s}}) -  g_\lambda( \pmb{\mathscr{s}}') | < \delta
\end{align}
but $ g_\lambda( \pmb{\mathscr{s}}')  = e^{\lambda \tau} g_\lambda( \pmb{\mathscr{s}})$ by assumption hence
\begin{align}
| g_\lambda( \pmb{\mathscr{s}})| |1 - e^{\lambda \tau}| < \delta
\end{align}
which is a contradiction since $\tau$ can be made arbitrarily large and $\delta$ arbitrarily small. The non-existence of Koopman modes satisfying $g_\lambda(\pmb{s}(t+\tau)) = e^{\lambda \tau} g_\lambda(\pmb{s}(t))$ over all of state space is corroborated by numerical evidence \cite{Parker2019}. In so far as a turbulent attractor is mixing one does not expect a finite dimensional linear subspace for the Koopman modes (except for the constant observable). See \cite{HankelDMD} for a similar statement with regards to the Lorenz attractor.  We take the above proof as a plausible argument for the use of a piecewise discontinuous basis in the representation of Koopman modes.

For stochastic dynamical systems, one expects that the Koopman modes (the Koopman operator is defined as the adjoint of the Fokker-Planck operator in that context) are continuous functionals of the state but no longer obey the relation $g_\lambda(\pmb{s}(t+\tau)) = e^{\lambda \tau} g_\lambda(\pmb{s}(t))$. Heuristically this is because the noise in the dynamics acts as a diffusion in probability space, which smooths out non-smooth fields. Equation 6.55 of \cite{StochasticKoopman2020} is illuminating. The takeaway is that, for a stochastic differential equation 
\begin{align}
d\pmb{s} =  \pmb{U}(\pmb{s})dt + \epsilon d\pmb{W}
\end{align}
where $\epsilon$ is the noise variance and $d\pmb{W}$ is a $d-$dimensional Wiener process, the Koopman mode evolves according to 
\begin{align}
d g_{\lambda}(\pmb{s}(t)) &= \lambda g_{\lambda}(\pmb{s}(t)) dt + \epsilon \nabla g_{\lambda}(\pmb{s}(t)) d\pmb{W},
\end{align}
The composition with the state variable is necessary because $g_\lambda : \mathbb{R}^d \rightarrow \mathbb{R}$, but an important note is that one cannot consider the evolution of $g_\lambda$ independently from where it is being evaluated in state space.
In the limit that the noise goes to zero, $\epsilon \rightarrow 0$, the gradient term, $\nabla g_{\lambda}$, can go to infinity at particular points in state space, as would be expected in a two-well stochastic potential. Consequently, pathologies are unexpected in linear systems.

As another point, although they are often called Koopman modes in the context of PDEs, they should not be confused with spatial modes.  Koopman modes are eigenoperators (in analogy to eigenvectors and eigenfunctions in lower dimensional contexts), i.e., functionals that act on a state. On the other hand the right eigevenctors of the transfer operator do act as projection operators to Koopman modes. The following section goes through a concrete example, but we remain abstract here for the moment. 

A continuum of observables indexed by $\pmb{x}$ define statistical modes as 
\begin{align}
\label{projection_operator}
G_{\lambda}(\pmb{x}) \equiv \int_{\mathcal{M}} d \pmb{\mathscr{s}} g^{\pmb{x}}(\pmb{\mathscr{s}}) \mathcal{P}_{\lambda}( \pmb{\mathscr{s}} )
\end{align}
where $\mathcal{P}_{\lambda}$ is a right eigenvectors of the transfer operator, i.e. 
\begin{align}
\mathcal{T}^{\tau}\mathcal{P}_{\lambda} = e^{\lambda \tau} \mathcal{P}_{\lambda}.
\end{align}
Equation \ref{projection_operator} projects the part of the observable $ g^{\pmb{x}}$ onto the appropriate Koopman mode\footnote{Due to bi-orthogonality of left and right eigenvectors.}.  We consider $G_{\lambda}(\pmb{x})$ as a mode, although it is perhaps more appropriate to call it called a modal amplitude of the associated Koopman mode. Whether or not the set of Koopman modes form a complete basis so that $g^{\pmb{x}} =  \sum_{\lambda} G_\lambda (\pmb{x}) g_{\lambda}$ is unclear. The implication is that an arbitrary observable $g^{\pmb{x}}$ could be fundamentally unpredictable if it cannot be expressed as a sum of Koopman modes. 

In the next section, we discuss the numerical approximation to Koopman modes.

\subsection{Numerical approximation}
\label{Numerical_Koopman_Modes}
The numerical Koopman modes are the left eigenvectors of the matrix $Q$, denoted by $\pmb{g}_{\lambda}$ and their approximation as functionals acting on the state is given by
\begin{align}
g_\lambda(\pmb{\mathscr{s} }) \approx [\pmb{g}_\lambda]_{\mathcal{E}(\pmb{\mathscr{s} })},
\end{align}
where $[\pmb{g}_\lambda]_n$ is the $n'th$ component of the eigenvector $\pmb{g}_{\lambda}$. Hence we first apply the embedding function to the state and then use the integer label to pick out the component of the eigenvector $\pmb{g}$.

At each moment in time, we plot the approximate Koopman mode 
\begin{align}
g_\lambda(\pmb{s}(t) ) \approx [\pmb{g}_\lambda]_{\mathcal{E}(\pmb{s}(t) )},
\end{align}
where the component of the vector $\pmb{g}$ is given by $\mathcal{E}(\pmb{s}(t) )$. We show these dynamics for the first 30 simulated days of the Held-Suarez setup in Figure \ref{fig:held_suarez_koopman_modes}. Furthermore, we compute autocorrelations in two ways to check the fidelity of the numerical Koopman modes. This calculation is shown in Figure \ref{fig:held_suarez_koopman_modes}. We see that from the top right panel that the two methods of calculation agree for all times for mode 351 (red), but only for the first 5 or so days for Mode 6 (blue). Given the near-exponential structure of the decay using the time series, this suggests that there exists a perturbation to the existing generator that could align the time series and ensemble calculation. Perhaps the peak around 14 days is synonymous with the usual decorrelation time assumed for the atmosphere. The real component of the Koopman modes as a function of partition index are shown in the bottom two panels. We see that mode 351 picks up on unlikely partitions of state space whereas mode 6 is evenly distributed amongst all states. A binary classification algorithm would divide partitions according to the positive and negative values of the mode.

And finally we show the Koopman mode amplitudes associated with the surface temperature field in Figure \ref{fig:held_suarez_koopman_mode_amplitudes}. Thus the observable is 
\begin{align}
g^{\pmb{x}_s}(\pmb{\mathscr{s}}) = \mathcal{T}_{\pmb{x}_s}
\end{align}
where $\pmb{x}_s$ is a surface temperature and $\mathcal{T}_{\pmb{x}_s}$ is the temperature observable defined by Equation \ref{temperature_definition}.
The projections are computed using the discrete analog to \ref{projection_operator}, 
\begin{align}
\label{projection_operator_discrete}
G_{\lambda}(\pmb{x}_s) \approx \sum_n g^{\pmb{x}_s}(\pmb{\sigma}^n) [\vec{\mathcal{P}}_{\lambda}]_n
\end{align}
where $\vec{\mathcal{P}}_{\lambda}$ is the right eigenvector of the generator $Q$ associated with eigenvalue $\lambda$. Furthermore, $[\vec{\mathcal{P}}_{\lambda}]_n$ denotes the $n'th$ component of the eigenvector. We see that there are oscillatory modes associated with wavelike patterns.

\iffigure
\begin{figure}
\begin{center}
\includegraphics[width=1.0\textwidth]{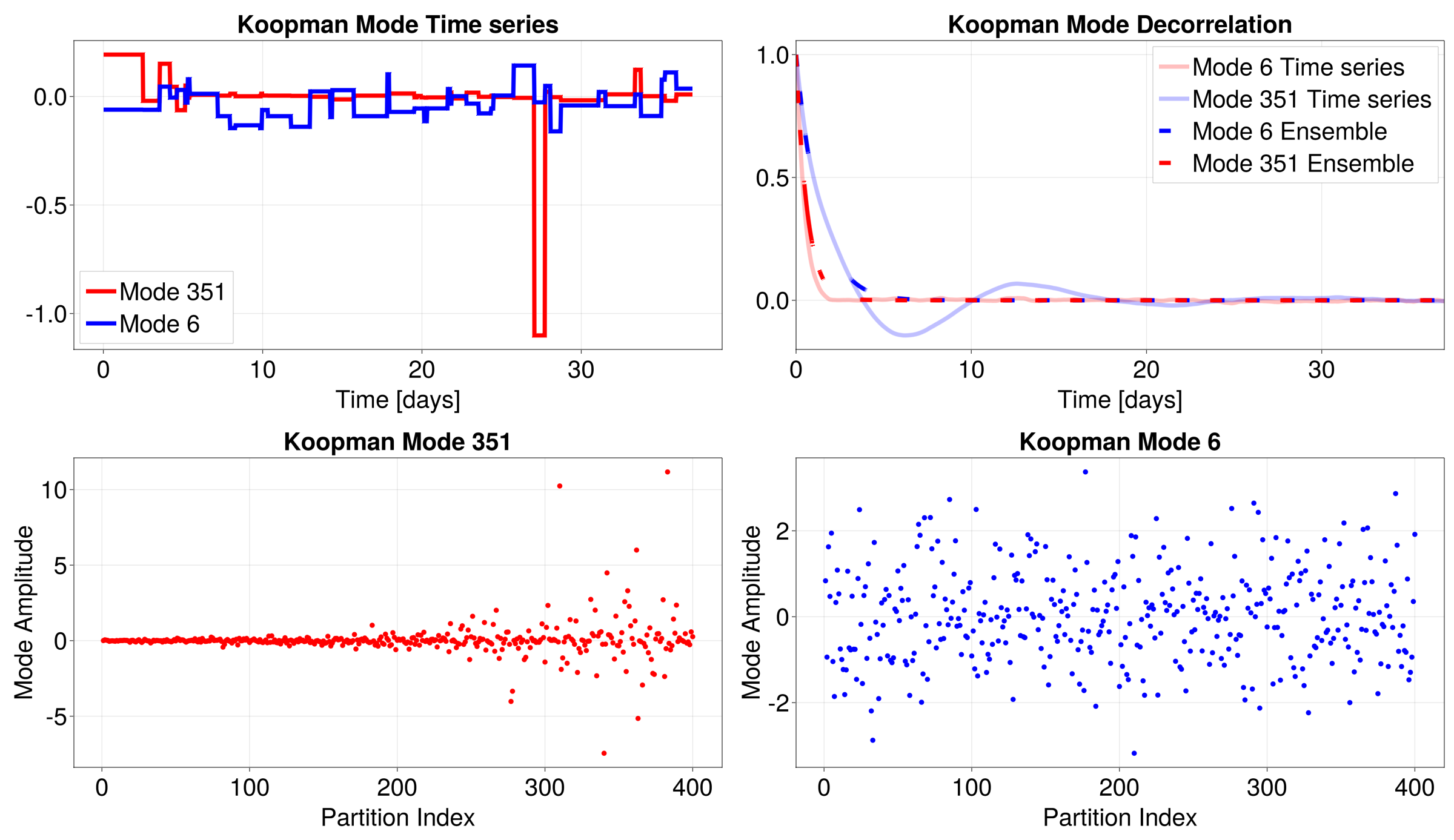}
\caption{Held-Suarez Koopman modes. We show two numerical Koopman modes in red and blue. The top left panel is the numerical Koopman modes as a function of time. The top right panel is the decorrelation timescales for the numerical Koopman modes computed by the generator (dashed) and the time series (solid). The bottom panels are the real part of the Koopman mode as a function of state index and thus implicitly as a functional of the state.}
\label{fig:held_suarez_koopman_modes}
\end{center}
\end{figure}
\fi

\iffigure
\begin{figure}
\begin{center}
\includegraphics[width=1.0\textwidth]{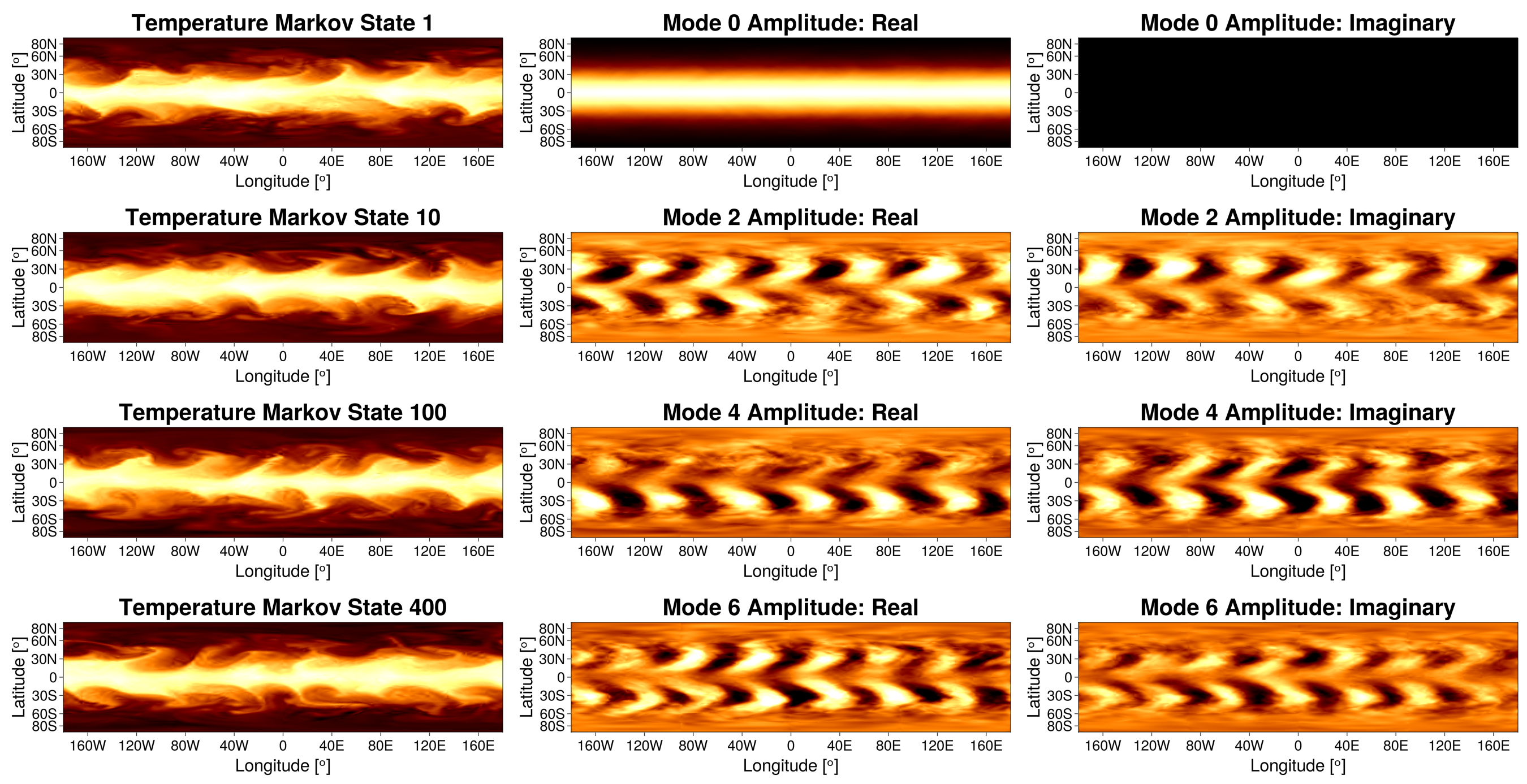}
\caption{Held-Suarez Koopman Mode Amplitudes. We show four representative surface temperature fields constructed from Markov states constructed from as well as their Koopman mode amplitudes by projecting the fields onto the real and imaginary parts of the associated Koopman mode. The statistically steady state is associated with mode zero. }
\label{fig:held_suarez_koopman_mode_amplitudes}
\end{center}
\end{figure}
\fi

\subsection{Matrix Decomposition for Decorrelations and Oscillations}
To further understand the timescales associated with the generator $Q$, we decompose the matrix into a negative semi-definite component and a component with purely imaginary eigenvalues. First, we assume that the generator $Q$ is ergodic so that it has one zero eigenvalue, and all other eigenvalues have strictly negative real parts.

Let $\vec{\mathbb{P}} = [\mathbb{P}(\Omega_1), ..., \mathbb{P}(\Omega_n)]$ be the normalized eigenvector corresponding to eigenvalue $\lambda = 0$, where we take the normalization to be 
\begin{align}
\pmb{1}^T \vec{\mathbb{P}} = 1
\end{align}
where $\pmb{1}$ is the vector of all $1's$. Under the ergodic assumption, all vector $\vec{p}$ entries are positive.

We split the matrix $Q$ into a negative semi-definite and pure imaginary part as follows
\begin{align}
\label{decomposition}
Q = \frac{1}{2}\underbrace{Q + P Q^T P^{-1}}_{\text{negative semi-definite}} + \frac{1}{2}\underbrace{Q - P Q^T P^{-1}}_{\text{imaginary eigenvalues}}
\end{align}
where $P = \text{Diagonal}(\vec{\mathbb{P}})$ is a diagonal matrix whose entries along the diagonal are the steady state distribution $\vec{\mathbb{P}}$. The relation $P^{-1} \vec{\mathbb{P}} = \pmb{1}$ holds. The proof that the matrix $Q + P Q^T P^{-1}$ is negative semi-definite is as follows. We first observe $R_E(g, dt) \leq R_E(g, 0)$ via the Cauchy-Schwarz inequality. More intuitively, this is just the statement ``observables eventually decorrelate". Then,
\begin{align}
R_T(g, dt) = \pmb{g}^T \exp(Q dt) P \pmb{g} \approx \pmb{g}^T  P \pmb{g} + dt \pmb{g}^T QP \pmb{g} \leq  \vec{g}^T  P \pmb{g} \Rightarrow   \pmb{g}^T QP \pmb{g}  \leq 0
\end{align}
Since $QP$ is negative semi-definite and we can rescale $\pmb{g}$ as $\pmb{h} = P^{-1/2} \pmb{g}$, the matrix $\tilde{Q} = P^{1/2} Q P^{-1/2}$ is negative semi-definite. Thus the symmetric part $\tilde{Q}$ is negative semi-definite. Noting the similarity transformations
\begin{align}
Q + P Q^T P^{-1} &= P^{1/2} \left[ P^{-1/2}Q P^{1/2} + (P^{-1/2}Q P^{1/2})^T \right] P^{-1/2}
\\
Q - P Q^T P^{-1} &= P^{1/2} \left[ P^{-1/2}Q P^{1/2} - (P^{-1/2}Q P^{1/2})^T \right] P^{-1/2}
\end{align}
completes the proof since similar matrices have equivalent eigenvalues. Heuristically, the $Q + P Q^T P^{-1} $ part of the decomposition contributes to decorrelation in time and $Q + P Q^T P^{-1} $ contributes to oscillations in time.

Regarding the Held-Suarez generator, the timescales associated with the eigenvalues of $Q + P Q^T P^{-1}$ range from $\approx 0.25$ days to $\approx 1.6$ days, whereas the timescales associated with eigenvalues of $Q - P Q^T P^{-1}$ range from 25 days to 44 years. The purely oscillatory timescales are never observed due to the interaction between the two matrix components when evolving in time. We comment that the decomposition is especially useful when the two matrix components commute. Furthermore, the $Q + P Q^T P^{-1}$ matrix contains the numerical dissipation associated with the scheme. 

The symmetric part of this matrix has been commented on before by \cite{Froyland2005} when defining a time-reversible Markov chain from an irreversible one. Furthermore, see \cite{Klus2020} for splitting the drift and diffusion terms.
\section{Symmetries}
\label{symmetries}

In Section \ref{lorenz_fixed_point_partition}, the symmetries of the Lorenz equations were not incorporated directly into the generator. We rectify this deficiency here and outline a method for incorporating symmetries. The Lorenz equations are invariant with respect to the transformation $(x,y,z) \mapsto (-x, -y, z)$. In so far as one chaotic attractor exists, this symmetry is expected to apply to chaotic trajectories. To incorporate this symmetry, we take two steps.

The first step is to verify that the Markov states also satisfy this symmetry. Since $\pmb{\sigma}^1 \mapsto \pmb{\sigma}^3$, $\pmb{\sigma}^3 \mapsto \pmb{\sigma}^1$, and $\pmb{\sigma}^2 \mapsto \pmb{\sigma}^2$ under the symmetry operation, the Markov states, defined by the fixed points of the Lorenz equations, incorporate the symmetry. Generally, one must apply the symmetry operator to each Markov state and incorporate the ``symmetry states" as necessary.

The second step is to incorporate symmetries into the resulting Markov embedding. For example, in the case of the Lorenz equations, if we observe the sequence
\begin{align}
\text{first sequence} =  1, 1, 1, 2, 2, 3, 3, 1, 1
\end{align}
Then applying the symmetry operation to the above sequence yields 
\begin{align}
\text{second sequence} = 3, 3, 3, 2, 2, 1, 1, 3, 3
\end{align}
We then apply the Bayesian matrix construction on the first sequence and calculate the posterior distributions. We then use these posterior distributions as the new prior for a Bayesian matrix construction for the second sequence. Doing so yields a matrix that incorporates symmetry through data augmentation.

We show the expected values of the Lorenz Fixed Point Generator under this symmetry augmentation in Table \ref{moment_table_2}. We see that the expected values of quantities that should be zero are now zero. 

\begin{table}
\centering
\begin{center}
\begin{tabular}{ c c c c c c c c c c}
  &  $\langle x \rangle$ & $\langle y \rangle$ & $\langle z \rangle$ & $\langle xx \rangle$ & $\langle xy \rangle$ & $\langle xz \rangle$ & $\langle yy \rangle$ & $\langle yz \rangle$ & $\langle zz \rangle$ \\ 
 \hline
 ensemble & -0.0 & -0.0 & 23.8 & 63.5 & 63.5 & -0.0 & 63.5 & -0.0 & 642.4   \\
 \hline
   &   $\langle xxy \rangle$ & $\langle xxz \rangle$ & $\langle xyy \rangle$ & $\langle xyz \rangle$ & $\langle xzz \rangle$ & $\langle yyy \rangle$ & $\langle yyz \rangle$ & $\langle yzz \rangle$ & $\langle zzz \rangle$ \\
 \hline
 ensemble & 0.0 & 1713.2 & 0.0 & 1713.2 & 0.0 & 0.0 & 1713.2 & 0.0 & 17346.1 \\
\end{tabular}
\end{center}
\caption{Empirical Moments of the Lorenz Attractor. A comparison between ensemble averaging and time averaging.} 
\label{moment_table_2}
\end{table}

Similar considerations apply to other types of symmetries. For example, continuous symmetries are approximated as discrete symmetries, which can then use the methodology here.
\section{Held-Suarez}
\label{held_suarez}
Isaac Held and Max Suarez introduced a simplified atmospheric model test in \cite{HS_1994}. The test case purposefully did not specify dissipation mechanisms and was meant to be flexible as to which prognostic variables or coordinate systems were employed in its calculation. Its primary purpose was as a robust ``physics test" to be compared across different numerical schemes and equations of motion. In Section \ref{PDE_setup}, we specify the equations, and, in Section \ref{numerical_method}, the numerical discretization that was used. Finally, we conclude in Section \ref{further_hs_deetz} with a follow-up to some of the points made in Section \ref{applications} about holding times, the convergence of matrix entries, and eigenvalue sensitivities.

\subsection{Partial Differential Equation Setup}
\label{PDE_setup}

We choose to use an equation set that retains fully compressible dynamics and is formulated in terms of density, total energy, and Cartesian momentum as the prognostic variables, yielding the equations
\begin{align}
\partial_t \rho + \nabla \cdot \left( \rho \pmb{u} \right) 
&= 0
\\
\partial_t (\rho \pmb{u} ) + \nabla \cdot \left( \pmb{u} \otimes  \rho \pmb{u} + p \mathbb{I} \right) 
&= -\rho \nabla \Phi - 2 \left(\pmb{\Omega} \cdot \hat{r}\right)\hat{r} \times \rho \pmb{u}  - k_v \left( \mathbb{I} - \hat{r} \otimes \hat{r} \right) \rho \pmb{u} 
\\
\partial_t (\rho e ) + \nabla \cdot \left( \pmb{u} \left( p + \rho e \right) \right) 
&= - k_T \rho c_v \left(T - T_{\text{equilibrium}} \right)
\end{align}
where $\Phi = 2 G M_P r_{\text{planet}}^{-1} - G M_P  r^{-1} $ is the geopotential, $\pmb{\Omega} = \Omega \hat{\bf z}$ is the planetary angular velocity, $\hat{\bf z}$ is the direction of the planetary axis of rotation, and $r$ is the radial direction in spherical coordinates. The Coriolis force is projected to the radial component so that small planet analogs (which we use for the simulation in Section \ref{applications}) have a climatology similar to Earth. Furthermore, the variable $T_{\text{equilibrium}}$ is the radiative equilibrium temperature depending on latitude ($\varphi$) and pressure $\sigma = p/p_0$,
\begin{align}
    T_{\text{equilibrium }} (\varphi, \sigma) &= \text{max}\left(T_{\text{min}}, [T_{\text{equator}} - \Delta T_y \sin^2(\varphi) - \Delta \theta_z \ln(\sigma) \cos (\varphi) ] \sigma^{R_d / c_p} \right),
\end{align}
and the parameters $k_v$, $k_T$ are the inverse timescales for momentum damping and temperature relaxation, respectively, with
\begin{align}
    k_v &= k_f \Delta \sigma
    \quad \text{and}\quad
    k_T = k_a + (k_s - k_a) \Delta \sigma \cos^4(\varphi),
\end{align}
with $\Delta \sigma = \text{max}\left\{ 0, (\sigma - \sigma_b)/(1 - \sigma_b) \right\}$.
The temperature and pressure are,
\begin{align}
    T &= \frac{1}{c_v \rho} \left( \rho e - \frac{1}{2} \rho  \| \pmb{u} \|^2 - \rho \Phi \right) \quad
    \text{and} \quad
    p = \rho R_d T.
\end{align}
The parameter values for the simulation setup are in Table \ref{tab:hs_parameters}.

\begin{table}
\begin{tabular}{ |c|c|c|c|} 
\hline
\text{parameter} & \text{value} & \text{unit} & \text{description} 
\\
\hline
$\mathcal{X}$ & 80 & - & scaling parameter
\\
$z_{top}$ & $3 \times 10^4$ & m & atmosphere height
\\
$r_{\text{planet}}$ & $6.371 \times 10^6 / \mathcal{X}$ & m & planetary radius
\\ 
$R_d$   & 287 & $
\text{m}^2 \text{ s}^{-2} \text{ K}^{-1}$ & gas constant for dry air
\\
$\Omega$ & $2 \pi / 86400 \times \mathcal{X}$ & $\text{s}^{-1}$ & \text{Coriolis magnitude}
\\
$p_0$ & $1 \times 10^5$ & $\text{kg } \text{m}^{-1} \text{ s}^{-2}$ & reference sea-level pressure 
\\ 
$T_{min}$ & 200 & K & minimum equilibrium temperature 
\\ 
$T_{equator}$ & 315 & K & equatorial equilibrium temperature 
\\ 
$\sigma_b $ & 0.7  & - & dimensionless damping height 
\\
$c_v$ & 717.5 & $\text{J} \text{ kg}^{-1} \text{ K}^{-1}$ & specific heat capacity of dry air at constant volume
\\
$c_p$ & 1004.5 & $\text{J} \text{ kg}^{-1} \text{ K}^{-1}$ & specific heat capacity of dry air at constant pressure 
\\
$k_f$ & $ \mathcal{X} / 86400 $ & $\text{s}^{-1}$ & damping scale for momentum
\\
$k_a$ & $\mathcal{X}  / (40 \times 86400) $ & $\text{s}^{-1}$ & polar relaxation scale
\\
$k_s$ & $\mathcal{X} / (4  \times 86400) $ & $\text{s}^{-1}$ & equatorial relaxation scale
\\
$\Delta T_y$ & 60  & $\text{K}$ & latitudinal temperature difference
\\ 
$\Delta \theta_z$ & 10 & $\text{K}$ & vertical temperature difference
\\ 
$G$ & $6.67408 \times 10^{-11}$ & $\text{kg}^{-1} \text{ m}^3 \text{ s}^{-2}$ & gravitational constant 
\\ 
$M_P$ &  $5.9722 / \mathcal{X}^2 \times 10^{24}$ & kg  & planetary mass
\\
\hline
\end{tabular}
\caption{\label{tab:hs_parameters} Parameter values for the Held-Suarez test case. The value $\mathcal{X}=1$ corresponds to the standard test case, and $\mathcal{X}=80$ is the version that we use here.}
\end{table}

We use no-flux boundary conditions for density and total energy, free-slip boundary conditions for the horizontal momenta, and no-penetration boundary conditions for the vertical momentum. The initial condition is a fluid that starts from rest, $\rho \pmb{u} = 0$, in an isothermal atmosphere, 
\begin{align}
\label{held_suarez_ic}
p(r) = p_0 \exp \left(- \frac{\Phi(r) - \Phi(r_{\text{planet}})}{R_d T_{I}} \right) \text{ and } \rho(r) = \frac{1}{R_d T_I} p(r)
\end{align}
where we use $T_{I} = 285K$.

\subsection{Numerical Method} 
\label{numerical_method}
To approximate the equation of the previous section, we use the Flux-Differencing Discontinuous Galerkin method outlined in \cite{Souza_2022} and precisely formulated in \cite{Maciej2021}. We choose numerical fluxes that are Kinetic+Potential Energy preserving to help ensure the flow's nonlinear stability and Roe fluxes for dissipation. In addition, the low storage fourth order 14-stage Runge Kutta method of \cite{Niegemann2012}  is used for time stepping and induces a form of numerical dissipation. All simulations were run on an NVidia Titan V graphics processing unit.

The domain is a piecewise polynomial approximation to a thin spherical shell of radius $r_{\text{planet}}$ and height $z_{top}$. The thin spherical domain is partitioned into curved elements and uses an isoparametric representation of the domain, and the cubed sphere mapping by \cite{Ronchi_1996}. In essence, this choice represents the domain as a piecewise polynomial function where the order of the polynomial corresponds to the order of the discretization \cite{Gassner_2021}. The metric terms are treated as in \cite{Kopriva_2006} and satisfy the discrete property that the divergence of a constant vector field is zero, i.e., the metric terms are free-stream preserving.

We use 4 elements in the vertical direction, $6 \times 6^2$ elements for the sphere's surface ($6^2$ elements per cubed sphere panel), and order 6 polynomials within each element. Given that we have 5 prognostic states (density, the three components of the Cartesian momenta, and total energy), this leads to a total of $5 \times 4 \times (6 \times 6^2) \times 7^3 = 1,481,760$ degrees of freedom—the horizontal acoustic CFL limits timesteps.

\subsection{Partition Properties and Uncertainty Quantification}
\label{further_hs_deetz}

We show two figures for investigating convergence. In Figure \ref{fig:holding_time_convergence}, we show the generator's inverse holding times (diagonal entries) for the first 16 most probable states. We see that there appears to be convergence to the matrix entries over disparate time intervals. 

\iffigure
\begin{figure}
\begin{center}
\includegraphics[width=1.0\textwidth]{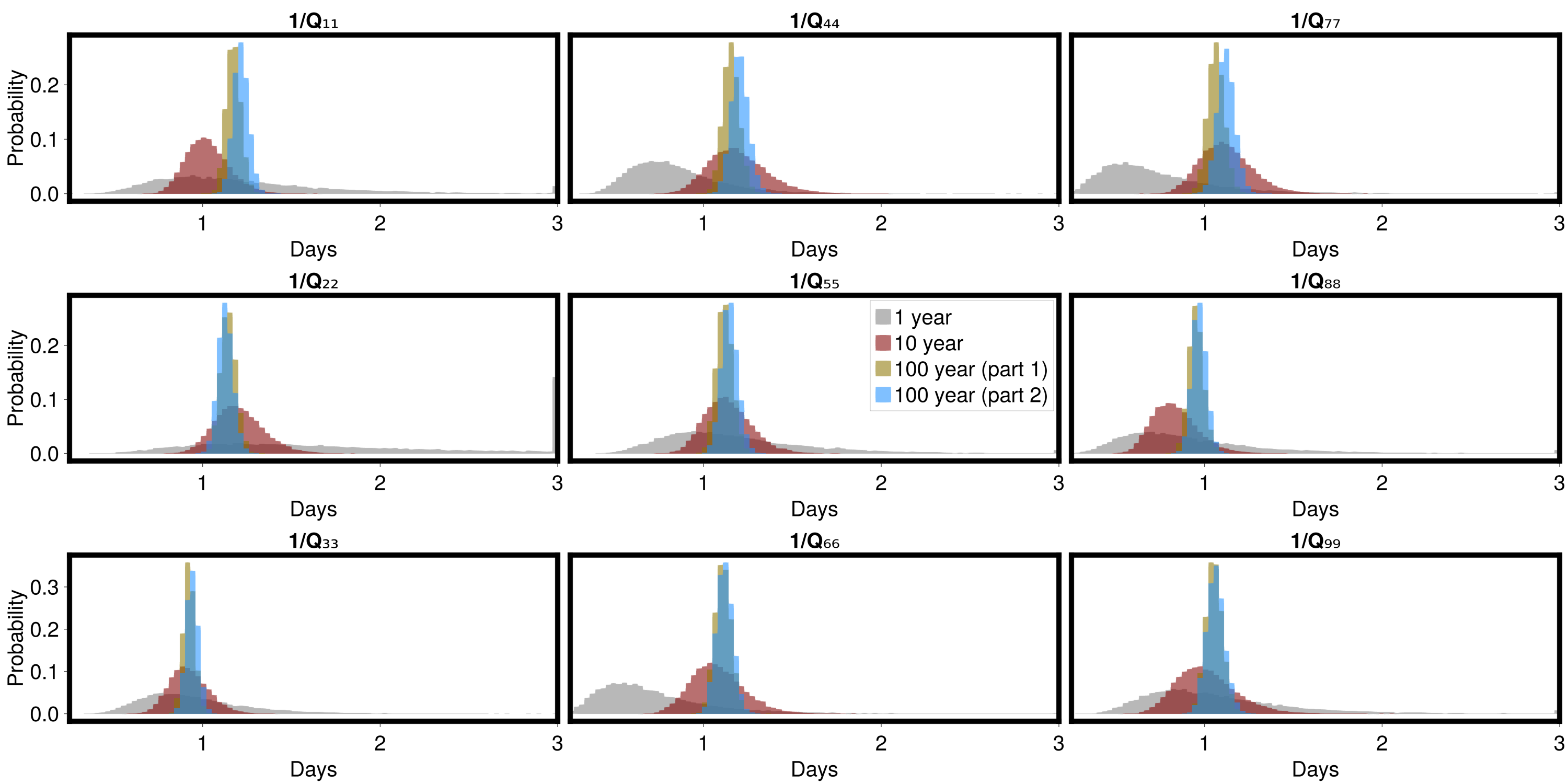}
\caption{Held-Suarez Generator Rate Entries. The uncertainty with respect to the inverse rates is shown for various time intervals. In grey, red, blue, and gold, we show the uncertainty corresponding to 1 year, 10 years, 100 years, and another (separate) 100-year simulation. We see that there is a significant overlap in the two 100-year estimates. }
\label{fig:holding_time_convergence}
\end{center}
\end{figure}
\fi

In Figure \ref{fig:held_suarez_decorrelation_timescales}, we show the real part of the inverse eigenvalue as distributions from random samples of the generator matrix. This variable corresponds to the decorrelation time scale as given by the partition. The Bayesian approach suggests that we cannot trust the slowest decorrelation scale obtained from the numerical solution since it varies between 1.5 days to 20 days. On the other hand, the other eigenvalues cannot be dismissed as meaningless since the probability distributions overlap with one another over data collected over disjoint subsets of time. As a technical note, uncertainty propagation of the eigenvalues can potentially be accelerated by using the eigenvalue decomposition of the mean generator as a guess for an iterative procedure.

\iffigure
\begin{figure}
\begin{center}
\includegraphics[width=1.0\textwidth]{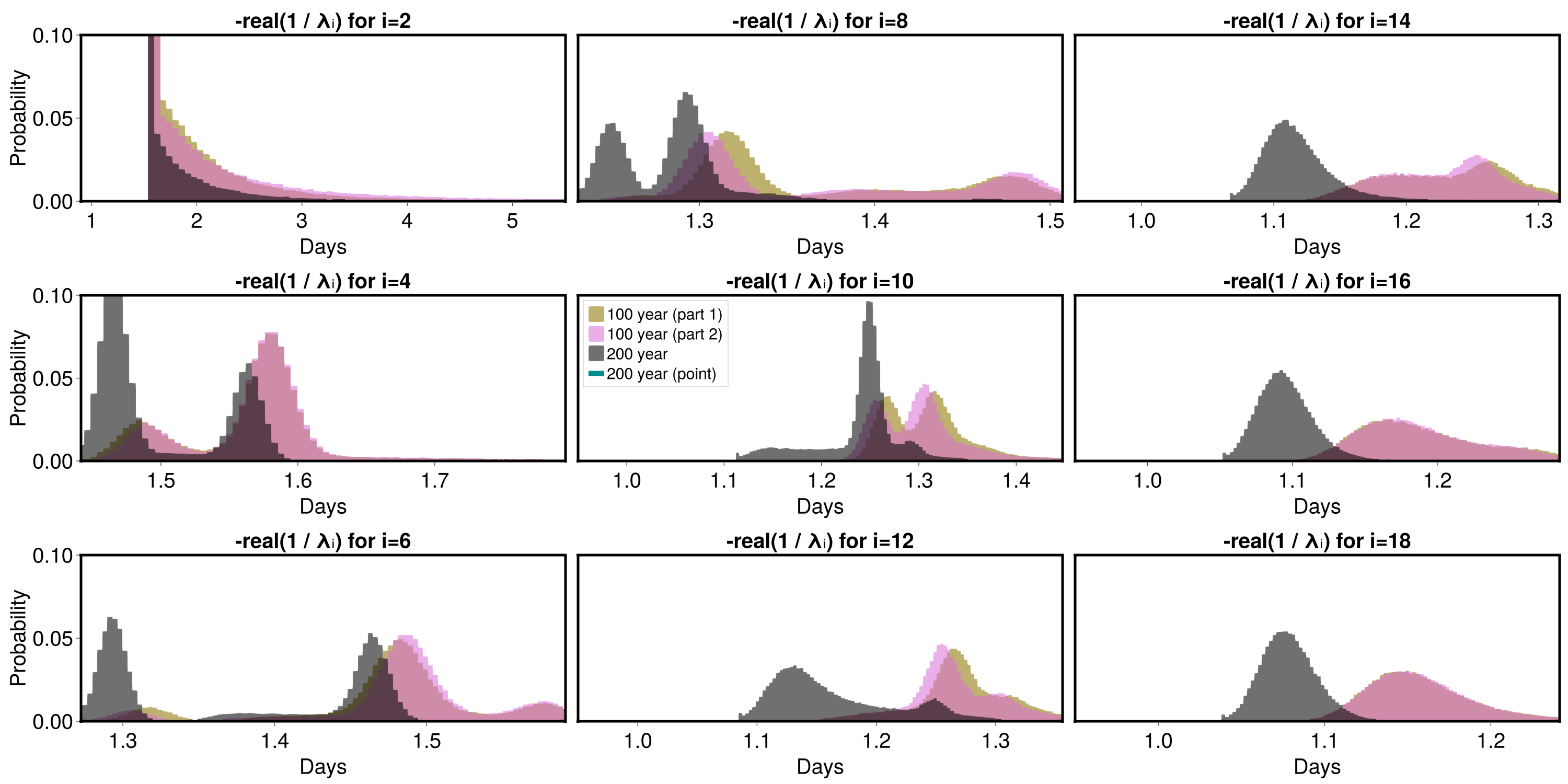}
\caption{Held-Suarez Generator Decorrelation Timescales. We propagate the uncertainty with respect to the random generator to look at the inferred distribution of eigenvalues. We propagate uncertainty for three cases: the entire 200 years, the first 100 years, and the second 100 years. Furthermore, we display the point estimate of the 200-year generator as calculated from the mean of the random matrix. The slowest decorrelation timescale is the most sensitive to perturbations, and the other eigenvalues are less so. }
\label{fig:held_suarez_decorrelation_timescales}
\end{center}
\end{figure}
\fi

As was done in Section \ref{lorenz_fixed_point_partition} we show the holding time distributions for particular states. Given that we have four hundred states, we show the holding times for the first three most probable states in Figure \ref{fig:held_suarez_holding_times}. Quantiles are approximately exponentially distributed but become imperfect upon closer inspection, as expected.

\iffigure
\begin{figure}
\begin{center}
\includegraphics[width=1.0\textwidth]{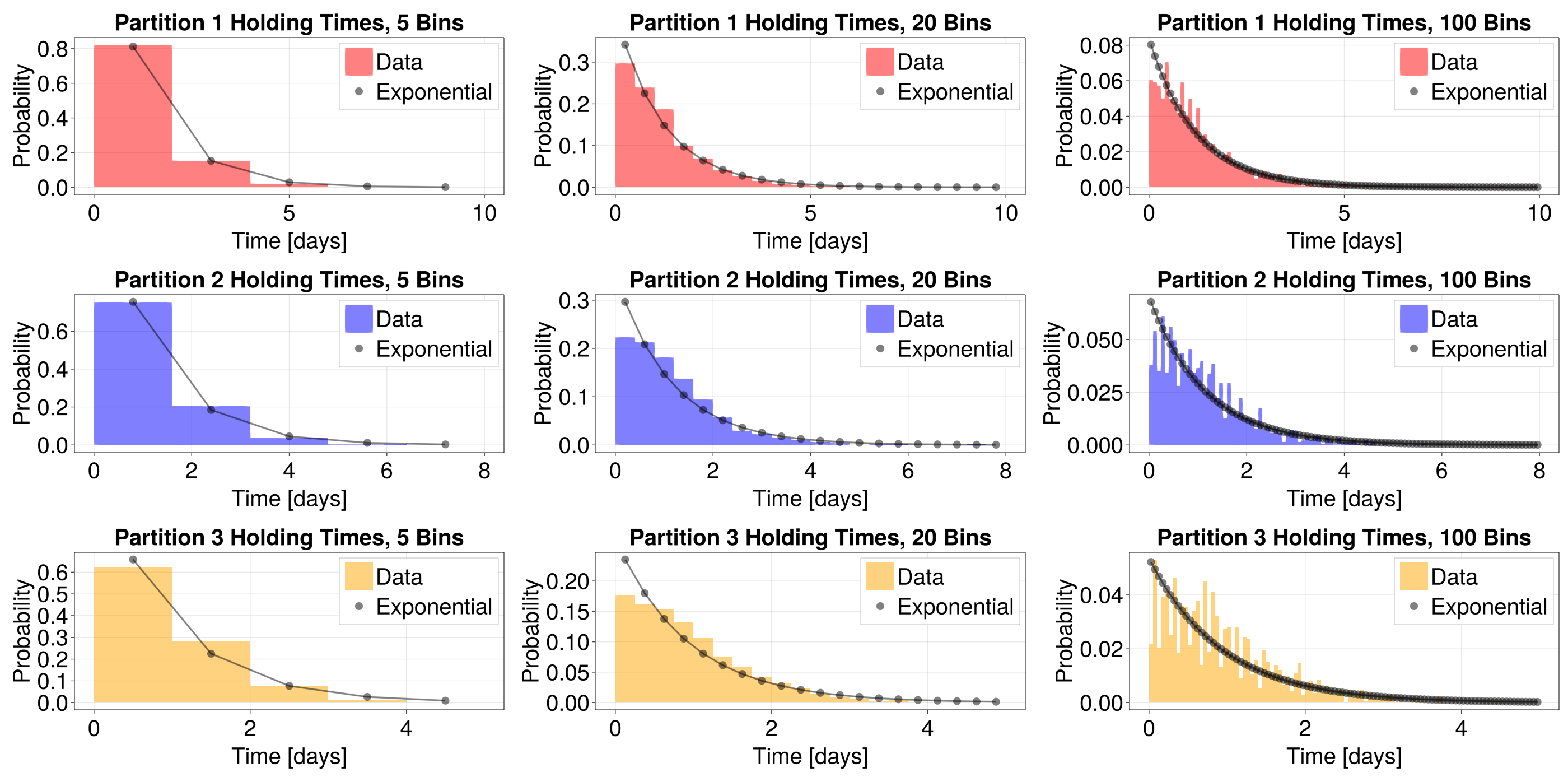}
\caption{Held Suarez Holding Times: Continuous Time Markov Model vs Timeseries Empirical Distribution. The amount of time spent in a state is approximately exponentially distributed. From the range of possible times, we see that state 3 has the largest holding time. On the other hand, state 1 is the most probable and not as exponentially distributed.}
\label{fig:held_suarez_holding_times}
\end{center}
\end{figure}
\fi

\bibliographystyle{jfm}
\bibliography{references}


\end{document}